# Anomalous properties of spark plasma sintered boron nitride solids

Abhijit Biswas,[a,#,*] Peter Serles,[b,#] Gustavo A. Alvarez,[c] Jesse Schimpf,[d] Michel Hache,[e] Jonathan Kong,[e] Pedro Guerra Demingos,[e] Bo Yuan,[f] Tymofii S. Pieshkov,[a,g] Chenxi Li,[a] Anand B. Puthirath,[a] Bin Gao,[h] Tia Gray,[a] Xiang Zhang,[a] Jishnu Murukeshan,[a] Robert Vajtai,[a] Pengcheng Dai,[h] Chandra Veer Singh,[e] Jane Howe,[e] Yu Zou,[e] Lane W. Martin,[a,d,i] James Patrick Clancy,[f,*] Zhiting Tian,[c,*] Tobin Filleter,[b,*] & Pulickel M. Ajayan[a,*]

**AFFILIATIONS**

[a]Department of Materials Science and Nanoengineering, Rice University, Houston, TX, 77005, USA

[b]Department of Mechanical & Industrial Engineering, University of Toronto, 5 King's College Road, Toronto, M5S 3G8, Canada

[c]Sibley School of Mechanical and Aerospace Engineering, Cornell University, Ithaca, NY 14853, USA

[d]Department of Materials Science and Engineering, University of California, Berkeley, 94720, USA

[e]Department of Materials Science and Engineering, University of Toronto, 184 College St, Toronto, M5S 3E4, Canada

[f]Department of Physics and Astronomy, McMaster University, Hamilton, L8S 4M1, Canada

[g]Applied Physics Graduate Program, Smalley-Curl Institute, Rice University, Houston, TX, 77005, USA

[h]Department of Physics and Astronomy, Rice University, Houston, TX, 77005, USA

[i]Materials Sciences Division, Lawrence Berkeley National Laboratory, Berkeley, 94720, USA

[#]Abhijit Biswas and Peter Serles equally contributed to this work

[*]Corresponding Authors: **abhijit.biswas@rice.edu, zhiting@cornell.edu, clancyp@mcmaster.ca, filleter@mie.utoronto.ca, ajayan@rice.edu**

## Graphical abstract

We synthesized near-theoretically dense bulk crystalline hexagonal boron nitride (h-BN) ceramics by using high-temperature spark plasma sintering of h-BN powders. The spark-plasma sintered h-BN ceramics exhibit anomalous properties, unveiling the design of scalable lightweight structural materials, pivotal for next-generation technology.

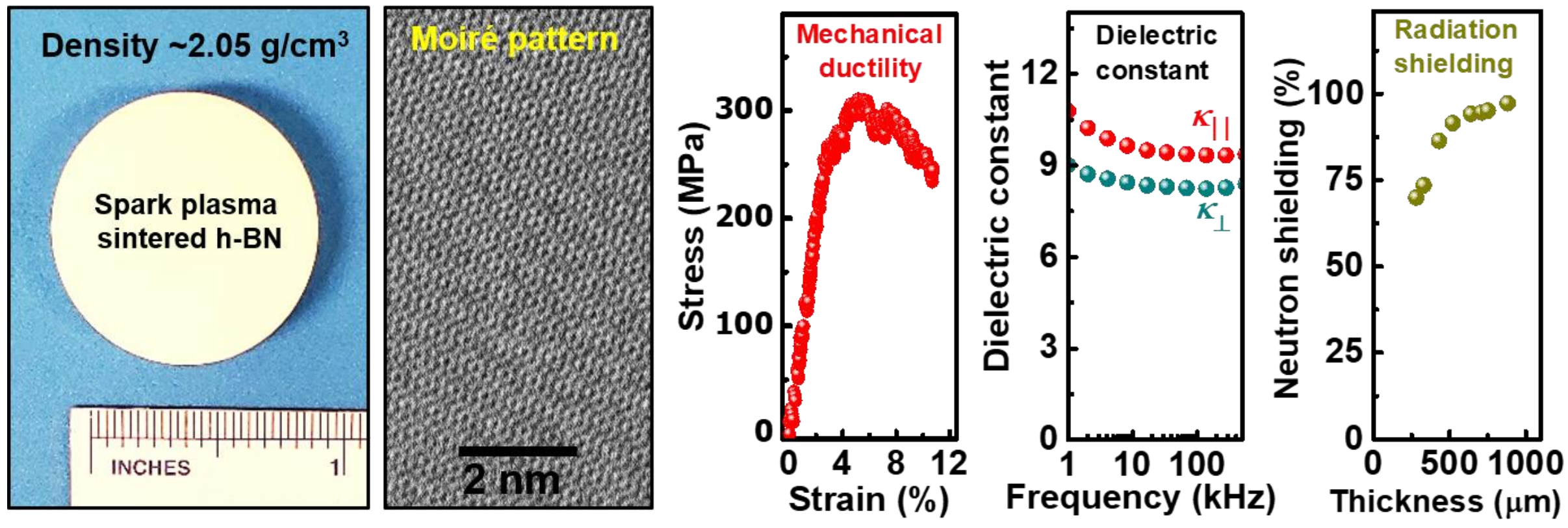

**Abstract**

Hexagonal boron nitride (h-BN) is a brittle ceramic with a layered structure, however, recent experiments have suggested that inter-layer structural engineering could be key to new structural and functional properties. Here we report the scalable bulk synthesis of high-density crystalline h-BN solids, by using high-temperature spark plasma sintering (SPS) of h-BN powders, which show high values of mechanical strength, ductility, dielectric constant, thermal conductivity, and exceptional neutron radiation shielding capability. Through exhaustive characterizations we reveal that SPS induces non-basal plane crystallinity, twisting of layers, and facilitates inter-grain fusion with a high degree of in-plane alignment across macroscale dimensions, resulting in near-theoretical density and improved properties. Our findings highlight the importance of material design, via new approaches such as layer twisting and interlayer interconnections, to create novel ceramics with properties that could go beyond their intrinsic limits.

## Introduction

Engineering of interatomic bonding and interfacial structures plays a pivotal role in tailoring the properties of layered materials, enabling applications that include structural, mechanical, optical, and electronics [1,2]. A recent study shows the excellent mechanical properties of hexagonal boron nitride (h-BN) solids produced through spark plasma sintering [3]. These properties have been attributed to the twisted stacked layers of h-BN and three-dimensional interconnected domains of these lateral stacks. In principle, the distorted or twisted lattices can have unique structural attributes that restrict movement of dislocations and lead to unique mechanical behavior [3]. Similarly, such interfacial modifications can also affect other properties such as scattering and phonon behavior [4,5]. Thus, engineering structure-property correlations especially in bulk assembled materials from low dimensional constituents such as two-dimensional (2D) structures requires fundamental understanding of nanoscale interfacial relationships (e.g. twists) and a judicious choice of processing is required to achieve this [3,6–8]. 2D h-BN forms a basal hexagonal lattice with primarily $sp^2$ hybridization, shows ultrawide-bandgap of ~5.9 eV, anisotropic thermal conductivity, and theoretically calculated static dielectric constant of ~3.76 (out-of-plane) and ~6.93 (in-plane) [8,9]. The theoretical density ($\rho$) of h-BN is ~2.1 g/cm$^3$ [10], but due to practical limitations, typical lab-synthesized bulk h-BN ceramics show a much lower packing density, ~60-70% of the theoretical value [11].

In the pursuit of high-density pristine bulk h-BN, various methods, such as high-temperature high-pressure (HTHP) sintering, mixing of two-grades of h-BN powders, using additives (e.g. impurities, metals, and other ceramics), as well as incorporation of the $sp^3$ bonded 3D cubic BN (c-BN) particles have been employed to produce h-BN with density modulations [12–17]. However, these composite materials may compromise on the high-performance properties inherent to h-BN. Therefore, achieving near-theoretical density and emergent functional properties in pristine bulk h-BN without any fillers or secondary phases using HTHP synthesis process is crucial for both atomic-scale engineering and its large-area scalability. For bulk pristine materials, HPHT sintering [18] can provide the energy necessary for the structural modifications (e.g. lattice distortions, twist between the layers) and consequent observation of functional properties, not seen by conventional synthesis process [3,4].

Here, we used spark plasma sintering (SPS) in pristine bulk h-BN without additives or fillers and obtained pure phase h-BN ceramic with achieved near-theoretical density. This is attributed to intergranular crystallization and reorientation of non-basal planes with twisted interfaces. Remarkably, the dense h-BN ceramic shows high mechanical hardness, deformability, yield strength, Young's modulus, anomalous static dielectric constant beyond theoretical limit, excellent thermal conductivity, and exceptional neutron radiation shielding capabilities. These observations hold great potential for bulk high-density ceramics of light-weight materials for numerous contemporary applications.

## Materials and methods

### ***Spark plasma sintering (SPS) of hexagonal BN powder***

We used commercially available high-purity (99.9% metal basis) h-BN powder, purchased from MSE suppliers, USA. The spark plasma sintering (SPS) was carried out on an SPS 25-10 machine (Thermal Technology LLC, California USA) at a constant uniaxial pressing pressure of 90 MPa and heating rate of 50 °C/min (at SPS facility in Texas A & M University, USA). The sintering temperature was kept at 1600/1700 °C. Sintering was carried out according to the following scheme: several grams of h-BN powder was placed in a graphite mold (diameter of 25 mm) and then placed in the sintering chamber under an initial pressure of 5 MPa. It was held at $\sim 2\times10^{-5}$ Torr for ~30 min, and then sintered for 60 min under atmospheric pressure of UHP (~99.999%) Argon gas medium. Sintering pressure started to ramp up to 45/90 MPa at the ramp rate of 3 MPa/minute. It takes ~30 minutes to get the pressure of 90 MPa (both temperature and pressure ramped up at the same time and they attained their max. values almost at the same time. The temperature of the SPS process was controlled by an optical pyrometer Raytek D-13127 (Berlin, Germany). After the SPS, pressure was released slowly at ~5 MPa/min, while the temperature was ramped down at ~100 °C/min. For conventionally sintered h-BN disk, h-BN powder was grounded in an agate mortar and pestle for ~30 min by adding a few drops of polyvinyl alcohol (PVA) binder. It was then pressed (with 4 Ton Load) to make a compact one-inch diameter disk. The sintered one-inch diameter pellet was then sealed inside a quartz tube in vacuum and sintered at 1000 °C for 12 hrs.

***Structural, spectroscopic, chemical, and microscopic characterizations (XRD, XPS, VBS, FESEM, HRTEM, FTIR, EBSD mapping, Raman spectroscopy, and REELS)***

X-ray diffraction (XRD) measurements of sintered disks were conducted with the Rigaku SmartLab X-ray diffractometer (Tokyo, Japan). The XRD measurements were carried out at 40 kV and 40 mA, utilizing a monochromatic Cu $K_{\alpha}$ radiation source (wavelength of 1.5406 Å) and a scanning rate of 1°/minute. Rietveld refinements of the XRD data were performed using the GSAS-II software package [19]. Temperature-dependent XRD measurements were performed on a Rigaku SmartLab system equipped with a Hypix 3000 detector. All measurements were done in the atmospheric environment. The temperature was raised at 10 K/min to targets, and was hold stable at each step for 5 minutes before the XRD measurements. For high-T XRD, few mg powders were scratched from the disk and grounded for 10 min to obtain fine grain.

X-ray photoelectron spectroscopy (XPS) was executed using the PHI Quantera SXM scanning X-ray microprobe, employing a monochromatic Al $K_{\alpha}$ X-ray source with a beam energy of 1486.6 eV. High-resolution core-level B, and N 1s elemental scans and valence band spectra (VBS) were performed with a pass energy of 26 eV and 69 eV. The Thermo Scientific™ Nexsa G2 XPS system was used for REELS measurements. A total of 10 scans were executed at a pass energy of 20 eV, and for a dwell time of 50 ms per scan, with the source beam energy set at 500 V. Fourier-transform infrared spectroscopy (FTIR) was conducted using the Nicolet 380 FTIR spectrometer equipped with a single-crystal diamond window. Raman spectroscopy measurements were taken with the Renishaw inVia confocal microscope, utilizing a 532 nm laser as the excitation source.

Surface topography was analyzed using a field emission scanning electron microscope (FESEM) model FEI Quanta 400 ESEM FEG. To minimize the charging effects, a thin layer (~10 nm) of gold (Au) was sputtered onto the BN particle surface. For EBSD mapping, Thermo Fisher Apreo 2 SEM with an EDAX Velocity Plus EBSD Detector was used with an applied voltage of 10 kV and current of 1.6 nA. For a cross-sectional view of the disk, cuts were made in FEI Helios NanoLab 660 Dual Beam microscope. Samples were coated with Pt metal, for which we deposited a 1 μm layer of Pt using a gas injection system and electron/ion beam. The Pt layer serves the purpose of protecting the top layers of the sample and their interface. Further cross-section trench was cut using a focused ion beam (FIB) with ~5 μm depth. SEM imaging of the cross-section was done in secondary electrons mode with 5 and 10 kV voltages and 0.1, 0.2 nA currents. Energy-dispersive X-ray analysis (EDX) analysis was conducted at 10 kV using the Oxford EDX detector and AZtec software for analysis. To calculate the pores density from FESEM images, we applied intensity threshold masks. Images were filtered to remove the intensity gradient due to charging of the surface by dividing the original image on the duplicated one with a gaussian filter (Kernel value 25). Next, after adjusting the threshold values of each image they were binarized before conducting measurements of the respective areas. All image processing was done in ImageJ software. The accuracy of the method depends on the uniformity of intensity as some of the smaller pores were excluded from the calculations.

For HRTEM, powders from the SPS h-BN disk were dispersed into the ethanol solution and sonicated in an ultrasonic bath for 10 min. Then we put a few drops of solution onto the carbon-coated Cu-grid, and dried it for 48 hrs. The Cu-grid was then mounted into the HRTEM chamber and images were recorded using Titan Themis operating at 300 kV.

***Dielectric measurements***

For dielectric measurements, we mechanically polished a disk and thinned down it to a few hundred µm. Dielectric constant ($\kappa$) was calculated from measured capacitance ($C$) according to $C = \frac{\kappa \epsilon_0 A}{t}$, where $\epsilon_0$ is the permittivity of free space ($8.85 \times 10^{-12}$ F/m), $A$ is the capacitor (top contact) area, and $t$ is the thickness of the sample. Sample thickness of 225 µm was measured with an optical microscope at the edge nearest to the capacitor**.** All measurements were taken from parallel-plate capacitor devices. The bottom of the sample was coated in silver paint to serve as a bottom electrode. Top contacts were made with smaller dots of silver paint (~1×1 mm$^2$). Samples were mounted on a piece of AlN using silver paint to provide an electrically insulating but thermally conducting barrier between the sample and heater. The AlN was then mounted on a resistive Inconel heater using silver paint. Dielectric measurements were taken using a Keysight E4990A Impedance Analyzer. Capacitance and loss were measured as a function of frequency at a 1V oscillation amplitude. This was repeated as a function of temperature.

***Thermal conductivity measurements***

The thermal diffusivity of the one-inch diameter pellets was performed via the laser flash method (Linseis XFA 500 Xenon Flash Thermal Conductivity Analyzer). The laser flash technique implements a xenon flash, which heats the sample from one end by producing a programmed energy pulse. The temperature rise is determined at the rear surface with a high-speed infrared detector. The temperature rise curve, recorded over time, represents the variation in sample temperature induced by the activation of the xenon flash. Utilizing mathematical models and known parameters, it calculates thermal diffusivity, offering crucial insights into a material's heat conduction capabilities across diverse temperature ranges. A thin layer of graphite spray coating was applied to the surfaces of the pellet to promote laser absorbance. The measurements were conducted at room temperature with a 10 Joule/laser pulse. The density of the pellet was determined by the solid cylinder method.

The heat capacity measurements were obtained through TA Instruments Differential Scanning Calorimeter (DSC) Auto 2500. DSC consists of a single furnace where samples and the reference undergo a heat-cool-heat cycle under a controlled temperature program. The samples, encapsulated in an aluminum pan, along with an empty reference pan are placed on a thermoelectric disk surrounded by the furnace. As the furnace temperature is changed at a constant rate of 10 °C/min, heat is transferred to the sample and the reference; the differential heat flow is then measured by area thermocouples. In this study, the pellets were placed in hermetic aluminum pans, and the specific heat capacity was measured over the temperature range of 0-36 °C at a heating rate of 10 °C/min.

The amplitude of heat flow is the sum of a heat capacity component and kinetic component

$$q = C_p \, dT/dt + f(T,t)$$

where q is the sample heat flow, $C_p$ the sample specific heat capacity, dT/dt the heating rate, and f(T,t) the kinetic response at a specific temperature and time.

***Nanoindentation***

Nanoindentation measurements were performed on 25 locations across the sample using a Berkovich diamond indenter to 2µm depth (KLA Instruments, iMicro). Hardness and Modulus were calculated following the Oliver & Pharr method. Compression of the micro-pillars was also performed using nanoindentation with a 10µm diameter diamond flat punch tip. The stress was calculated based on the diameter of the pillar at half height and the strain based on the undeformed pillar height above the substrate. The yield strength was calculated as the maximum stress upon deviation from the linear elastic loading regime, and the failure strain was selected as the end of the ductile plateau.

### *Micro-Pillar Design*

The micropillars were prepared by focused ion beam milling (FIB, Hitachi NB 5000) with $Ga^+$ ion source. A 40-kV accelerating voltage was applied for coarse milling of the pillar to a diameter of 5 μm. A 20-kV followed by 10 kV accelerating voltage were used for final thinning of the pillar to reach 2 μm in diameter.

### *Neutron absorption*

Neutron absorption measurements were carried out using the McMaster Alignment Diffractometer (MAD) on Beamport #6 at the McMaster Nuclear Reactor (MNR). The MNR is an open pool reactor, currently operating at a power of 3 MW. Measurements were performed using a monochromatic neutron beam ($\lambda = 2.21$ Angstroms or $E = 16.75$ meV) with a flux of approximately $5\times10^5$ neutrons/cm$^2$/s at the sample position. Samples were mounted in transmission geometry ($2\theta = 0°$), and positioned behind a 3 mm diameter Cadmium pinhole. Absorption coefficients were determined by comparing the transmitted neutron intensity with and without sample in place. Transmitted neutron intensities were measured using a helium-3 proportional detector. A low efficiency U-235 fission counter was used as an upstream beam monitor to normalize the intensity of the incident beam.

### *Density Functional Theory*

Density Functional Theory (DFT) calculations were performed within the VASP software [20], with the PBE/GGA approximation, a plane-wave basis set, and PAW pseudopotentials. A cutoff of 550 eV was used in all calculations, and dispersion forces were included using Grimme's D3 method. A 2×2×1 supercell of h-BN with two layers was modelled with a 9×9×7 Gamma-centered k-point grid. Ionic relaxation was performed until forces were less than 0.01 eV/Å. For compression simulations, the typically used static approach was employed, in which the ions are relaxed for fixed values of applied compressive strain. Atomic partial charges were calculated with Bader charge analysis.

## Results and Discussion

### *Structural characterizations of SPS h-BN*

We employed SPS of micron scale h-BN powder to obtain a bulk ceramic with conformal geometry (**Fig. 1a**) and performed extensive structural and microscopic characterizations. The spark plasma sintered h-BN (**SPS h-BN**) shows all the relevant Bragg peaks in X-ray diffraction (XRD), which is further confirmed by the electron back-scattered diffraction (EBSD) inverse pole figure orientational mapping (**Figs. 1b** and **1c**). Raman spectroscopy shows in-plane $E_{2g}$ Raman peak with full-width at half maxima (FWHM) of $E_{2g}$ peak is ~12.58 $cm^{-1}$ (**Fig. 1d**), confirming excellent crystallinity of BN [21]. We also observed Fourier-transformed infrared (FTIR)-active transverse optical (TO) mode (**Fig. 1e**). Cross-sectional field-emission scanning electron microscopy (FESEM) shows the dense layer of h-BN with uniform distribution of B and N, and the particle view shows 2D sheets-like features with sizes ~1-3 μm (**Fig. 1f**). High-resolution transmission electron microscopy (HRTEM) shows the moiré-like patterns from SPS h-BN (**Fig. 1g** and Supporting information **Fig. S1**). The Fast Fourier Transform (FFT) image indeed shows two different oriented layers, with a twisted rotation angle of ~21.63°, 14.12°, and 8.99° (21.63° case is shown in inset of **Fig. 1g**) [3]. Furthermore, X-ray photoelectron spectroscopy (XPS) scans shows the presence of B−N bonding with π-Plasmons peaks (~9 eV apart from the main B-N peak), characteristic of h-BN (**Figs. 1h** and **1i**) [21]. Valence band spectroscopy (VBS) shows the s- and p-bands and the valence band maxima (VBM) of ~1.8 eV, below the Fermi level ($E_F$) (**Fig. 1j**). Finally, from reflection energy loss spectroscopy (REELS), we obtain the high bandgap value of SPS h-BN of ~5.85 eV (**Fig. 1k**), similar to the bulk h-BN [8].

The observation of Moiré for layered polycrystalline h-BN ceramic is remarkable as this has only been obtained till now for few-layer mechanically exfoliated twisted h-BN single crystals showing intriguing properties [22,23]. The HTHP SPS process possibly produces twisted layers through a combination of temperature, pressure, and rapid sintering dynamics. h-BN layers are held together by vdW forces. Under high-T conditions, these forces allow the layers to rotate relative to each other [3]. The applied pressure during SPS can induce shear stress between the layers, which can cause rotational misalignment, leading to the formation of twisted layers. Also, thermal expansion and contraction between layers during the rapid heating and cooling may also contribute in twist. All these factors facilitate the relative rotation and alignment of layers, resulting in a nanoscale twist even in bulk h-BN induced by the SPS process.

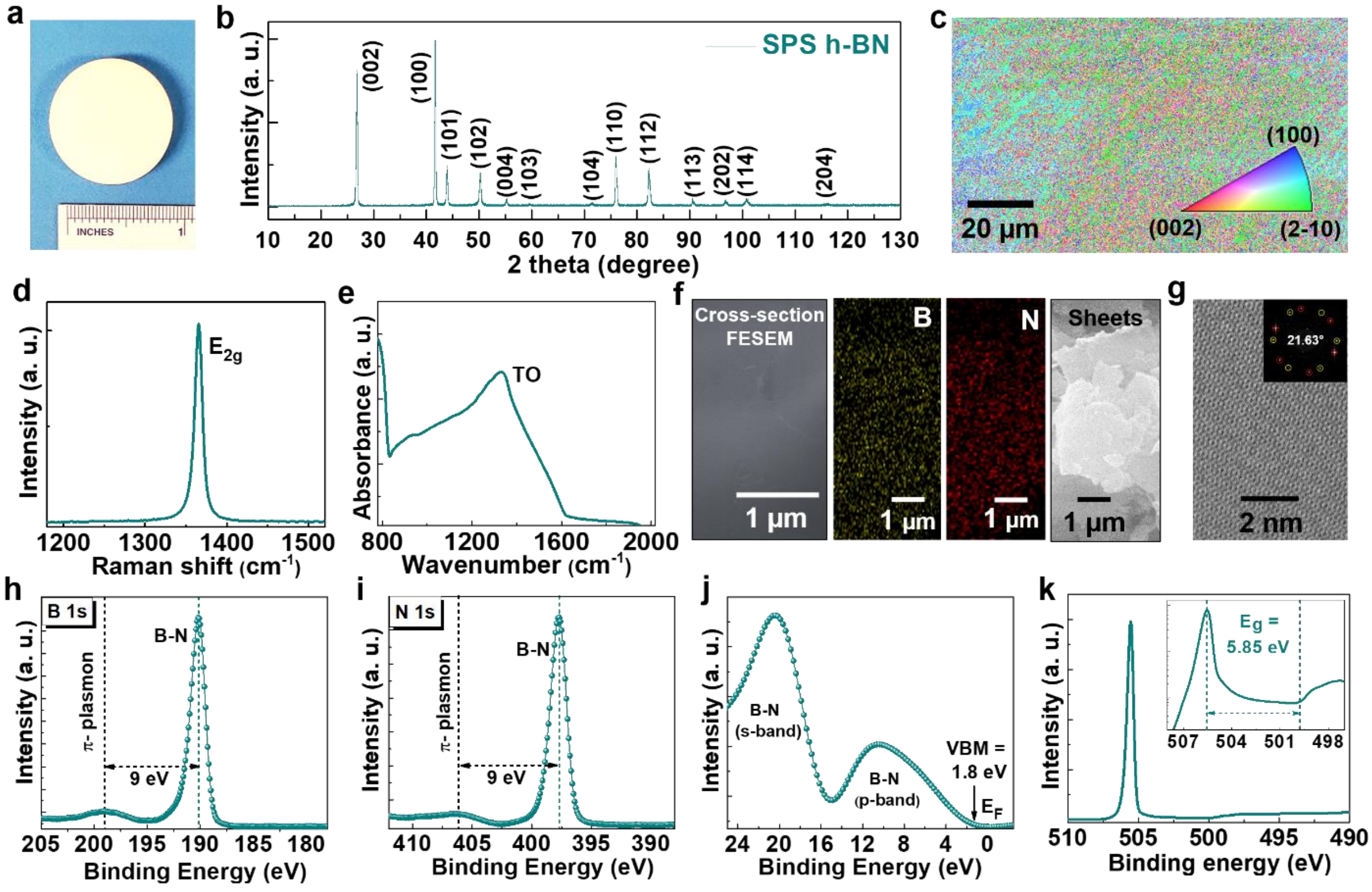

**FIG. 1 Structural characterizations of spark plasma sintered h-BN (SPS h-BN). (a)** SPS h-BN disk (one-inch in diameter) made from h-BN powders sintered at 1700 °C and 90 MPa, for 1 hr. **(b)** X-ray diffraction shows all the Bragg peaks correspond to h-BN. **(c)** EBSD inverse pole figure mapping shows the presence of various orientational planes for the SPS h-BN. **(d), (e)** Raman spectroscopy and FTIR spectra show in-plane $E_{2g}$ mode peaks with low full-width at half maxima, indicating excellent crystalline quality. **(f)** Cross-sectional FESEM image shows a denser layer for the SPS h-BN disk within uniform distribution of boron and nitrogen. Particle view of the disk shows the 2D-sheets like features. **(g)** HRTEM image showing moiré-like structures, and corresponding FFT image of two different oriented layers (inset), with a rotation angle between them (blue and red circular spots, respectively) of ~21.63°. **(h), (i)** Core-level XPS elemental scans shows the B-N bonding peaks with a π-Plasmon peak, characteristics of h-BN. **(j)** Valence band spectra (VBS) show that the valence band maxima (VBM) is ~1.8 eV below the Fermi level ($E_F$). **(k)** REELS show the band gap of ~5.85 eV.

We benchmark this SPS h-BN against a conventional lab-synthesis sintering process for h-BN powder (CS h-BN, see **method section** and **Fig. S2**). Structurally, we observe a few key changes in SPS h-BN. The intensity of the (002) Bragg peak is decreased, and shifted towards higher 2θ values, indicating the reduction in lattice *d*-spacing (**Fig. S3**). The SPS h-BN also demonstrates a drastic increase in other peaks as well [3]. We performed the Rietveld refinement of the XRD patterns and found that the SPS h-BN sample is compressed along the *c*-axis (-0.06%), and expanded within the ab-plane (+0.16%), providing a larger overall cell volume (+0.26%) (**Fig. S4**). HRTEM analysis and strain mapping of SPS h-BN further shows the presence of strain as well as in-plane lattice expansion (**Fig. S5**). Using the Scherrer equation, we found that average grain sizes increased to ≈35.0 nm as compared to ≈24.6 nm for the CS h-BN. This indicates that SPS promotes crystallization and reorientation along non-basal planes as well, despite the high surface energy of these orientations [24]. Moreover, the cross-sectional FESEM image of SPS h-BN demonstrates a highly dense material with minimal porosity at the micron scale while the CS h-BN has a significant number of voids and cracks (**Fig. S6**). From FESEM, we obtained minute pore densities (~1.1%) for SPS h-BN, as compared to pore densities of ~13.5% for CS h-BN. This further suggests the SPS promotes inter-grain fusions, in addition to the atomic recrystallization, as noted by the XRD and EBSD analysis.

The measured density of the SPS h-BN is $\rho$ = 2.05 $\pm$ 0.04 g/cm$^3$ which is 97.6% of the theoretical density of bulk 2D h-BN (~2.1 g/cm$^3$). This is especially notable in comparison to the CS h-BN which shows $\rho$ = 1.35 $\pm$ 0.03 g/cm$^3$, comparable to prior reports for typical lab-synthesized h-BN [11]. From XRD analysis, we found that the SPS h-BN displays strong preferred orientation, and the (001) orientation is prominently over-represented by 59.6%. This suggests that grains in the SPS h-BN sample must adopt a preferred orientation in order to reach such a high density. As such, we confirm the near-theoretical density resulting from the SPS process which promotes inter-crystallite fusion, minimizes porosity, and strains the atomic lattice thus creating a denser microstructure [18,25].

***Mechanical characterization of SPS h-BN***

The SPS treatment of h-BN creates a high-density material which is well suited to a range of industrial applications requiring robust mechanical performance and stability. Nanoindentation measurements of SPS h-BN indicate a drastic increase in Young's modulus ($E$) and Hardness ($H$) (**Figs. 2a** and **2b**), respectively. The SPS h-BN exhibits $E = 69.7 \pm 3$ GPa and $H = 283.3 \pm 9$ MPa, which is higher than or comparable with sintered h-BN and their composites from different synthesis methods [6,7,14,16,26–28]. These remarkable mechanical properties of SPS h-BN are especially stark in comparison to CS h-BN which exhibits minute mechanical stability with both low mechanical hardness and stiffness and is found to cleave easily by hand.

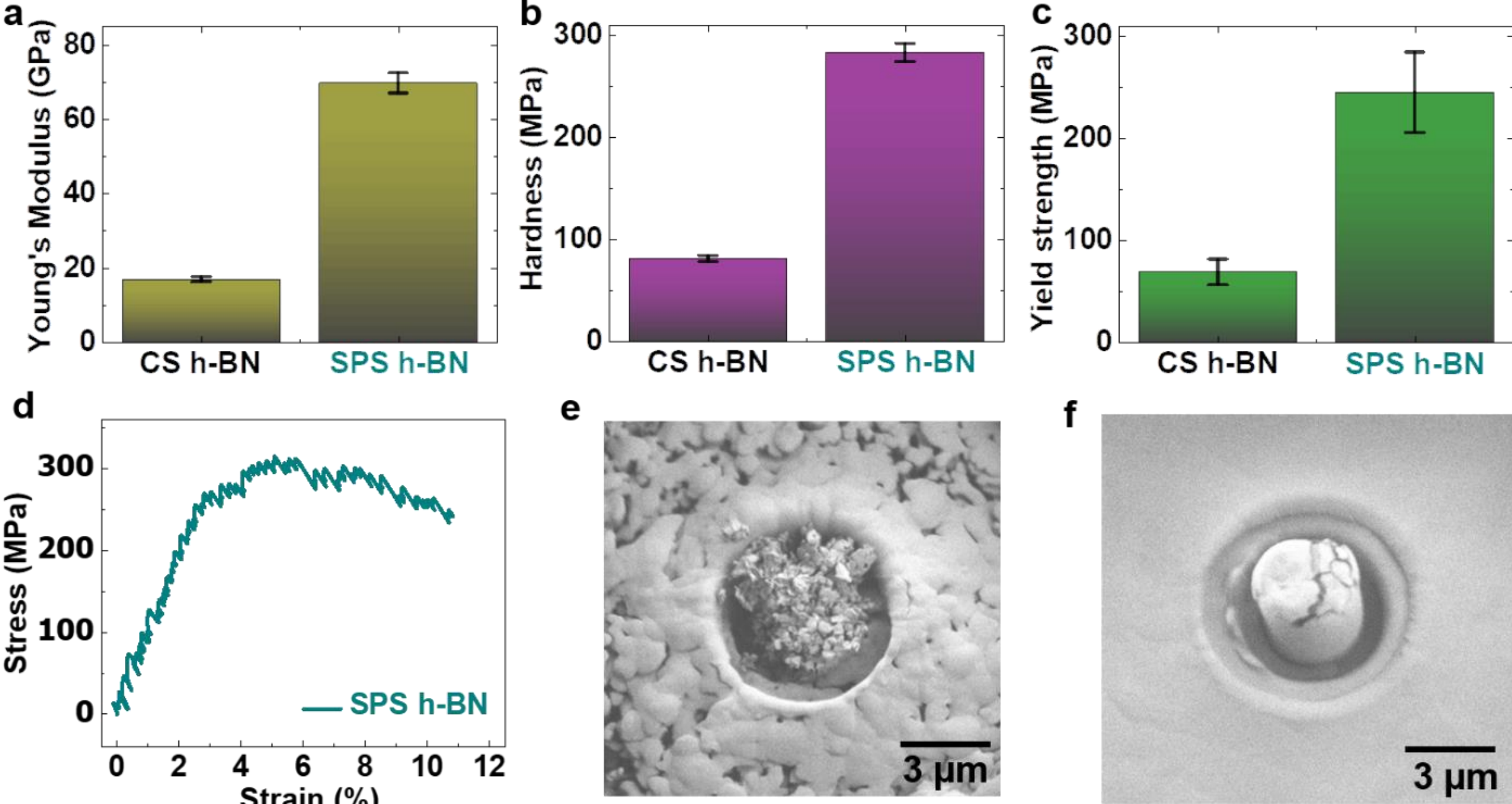

**FIG. 2 Mechanical properties of SPS h-BN**. **(a)** Young's modulus, **(b)** Hardness, and **(c)** Yield Strength of conventionally sintered h-BN (CS h-BN) and spark plasma sintered h-BN (SPS h-BN). As seen, SPS h-BN shows multi-fold increment in Young' modulus mechanical hardness, and yield strength. **(d)** Stress-strain failure test of SPS h-BN. A ductile yielding is noted at $\varepsilon = 3\%$ followed by a plateau until $\varepsilon = 11.5\%$ failure strain. **(e)** CS h-BN pillar crumbles under the mechanical loading and results in a powder-like morphology. **(f)** SPS h-BN pillar still remains intact under the mechanical loading but with evidence of crack propagation and shear fracture after failure.

The yield strength ($\sigma_y$) of these brittle ceramics is further shown by compressing focused ion beam (FIB)-cut micropillars (**Fig. 2c** and **Fig. S7**). The SPS h-BN exhibits a high yield strength of $\sigma_y = 245.1 \pm 39$ MPa in stark contrast to the conventionally sintered h-BN which offers a much lower yield strength.[3] Interestingly, rather than a brittle response typical of bulk ceramics and the CS h-BN [29,30], the SPS h-BN shows a ductile yield at $\varepsilon = 3\%$ followed by an elongated stress plateau up to strains of $\varepsilon$ =11.5% (**Fig. 2d**). This phenomenon is counter-intuitive for a high-density ceramic material, and has been attributed to twisted-layering of BN which offers high deformability due to microcrack suppression by localized delamination between interlocked nanoplatelets [3]. This effect allows for both high mechanical performance with incorporation of shock resilience and energy dissipation, a characteristic which is nonstandard for ceramics.

This failure mechanism can be further noted in the post-failure pillar morphology (**Figs. 2e** and **2f**). The CS h-BN micro-pillar crumbles under mechanical loading indicating a high degree of interparticle fracture due to limited bonding and high residual porosity during manufacturing, which is typical of nanocrystalline ceramics by condensed powders [31,32]. Conversely, even upon yielding of the SPS h-BN, the micropillar morphology is maintained without extensive crack propagation or shear fracture. This is in accord with the enhanced intercrystallite bonding and fusion noted in XRD and subsurface imaging, which results in a highly coherent and mechanically robust bulk ceramic material.

***Dielectric and thermal properties of SPS h-BN***

h-BN is widely considered to be the most promising gate insulator in 2D electronics [33]. For bulk h-BN, the calculated static dielectric constant in the out-of-plane direction ($\kappa_{\perp}$) is ~3.76, and the in-plane dielectric constant ($\kappa_{\parallel}$) is ~6.93 [9]. h-BN is free of dangling bonds and shows advantage over conventional insulators such as $SiO_2$ or $HfO_2$, which typically exhibit large densities of dangling bonds at the interfaces with 2D materials [34]. The bandgap of dielectrics is also important in achieving a favorable band offset with the target channel material. Typically, high bandgaps materials have relatively low-$\kappa$, whereas low bandgap materials show high-$\kappa$, but suffer from large leakage currents due to the lower bandgap and unfavorable band offset. Therefore, achieving higher $\kappa$ of SPS h-BN, would enable excellent gate control, essential for semiconductor devices.

Thus, considering the high theoretical density with compactness, we investigated the temperature and frequency-dependent static $\kappa$ and dielectric loss of bulk SPS h-BN. Surprisingly, for SPS h-BN, we obtain a much higher $\kappa_{\parallel}$ ~ 10.78, $\kappa_{\perp}$ ~9.01 (at 1 kHz) and $\kappa_{\parallel}$ ~9.35, $\kappa_{\perp}$ ~8.24 (at 100 kHz), which is beyond the theoretically calculated values (**Figs. 3a** and **3b**) [9]. In addition, we also have shown dielectric loss (inset of **Figs. 3a** and **3b**). There is little temperature dependence of the dielectric constant across the temperature range due to high thermal stability of h-BN. The measured $\kappa$ beyond the theoretically calculated values might be attributed to the charge induced from non-basal plane recrystallization, and twist between layers [3,22], which increases the ionicity due to a change in the local charge distribution. A small peak at ~475 K might be attributed to the random motion of nano-sheets across multiple measurements at elevated temperature. The magnitude of this change is too small to be associated with a phase transition [35]. We also examined the temperature dependent XRD of SPS h-BN, showing an absence of structural changes (h-BN is stable against temperature), except the thermal expansion related changes in 2θ-values (**Fig. S8**). Interestingly, for h-BN single crystals, it has been observed that twisted bi-layer material shows out-of-plane electric polarization [22,23]. We also investigated the polarization vs. electric field for our bulk SPS h-BN since it also shows twisted layers; however, we found the absence of hysteresis within the applied electric field of ±11 kV/cm (which may not nullify the possible presence of polarization at the nanoscale or at higher applied field) (**Fig. S9**).

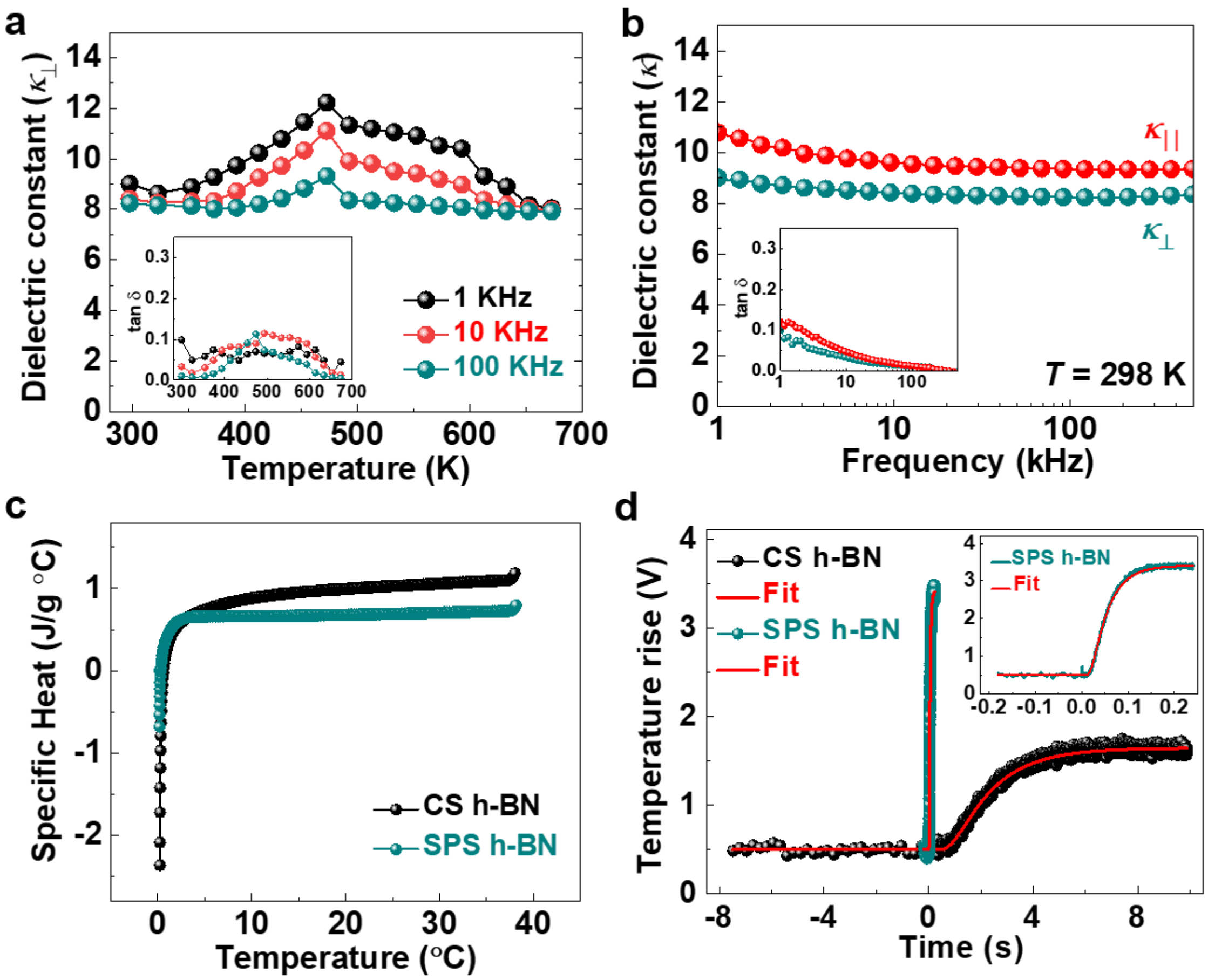

**FIG. 3 Dielectric and thermal properties of SPS h-BN**. **(a)** Temperature-dependent and **(b)** frequency-dependent static out-of-plane and in-plane dielectric constant of SPS h-BN. Inset shows the dielectric loss in respective cases. The dielectric constant is found to be much higher than the pristine bulk h-BN. The temperature and frequency dependency show negligible variation in the dielectric constant with low loss. A small hump in a temperature-dependent case is attributed to the randomness of sheets and ionic movements. **(c), (d),** Temperature-dependent specific heat capacity, and Dusza combined model fitting of the laser flash method temperature rise vs. time data, for conventionally sintered h-BN (CS h-BN) and spark plasma sintered h-BN (SPS h-BN). Inset shows the zoomed version of the temperature rise vs. time data plot for SPS h-BN. The SPS h-BN also shows a smoother temperature rise profile with less noise due to high density and less porosity.

The thermal conductivity ($K$) of h-BN is important for thermal management and energy-conversion of devices [36]. To determine the room temperature $K$ of h-BN, we measured the thermal diffusivity ($\alpha$) with the laser flash method, the specific heat capacity ($C_p$) with differential scanning calorimetry, the density ($\rho$) with the solid cylinder method, and finally implemented the relation $K = C_p \rho \alpha$. We obtained the temperature-dependent specific heat capacity ($C_p$) within 0-40 °C (**Fig. 3c**). Laser flash method data (temperature rise vs. time) were fitted with the Dusza combined model (**Fig. 3d**) [37]. For SPS h-BN, we obtained $\alpha = 13.1 \times 10^{-6} \pm 5.8 \times 10^{-8}$ m$^2$/s, $C_p$ = 696 $\pm$ 40.8 J/(kg-K), yielding a $K$ =18.72 $\pm$ 1.6 W/(mK) at room temperature, which is comparable with other fully dense ceramic materials.[38] In contrast, for CS h-BN, we obtained $K$ = 5.17 $\pm$ 0.6 W/(mK), comparable to the through-plane $K_z$ of h-BN (~2–5 W/mK) [39]. The laser flash method data for the SPS h-BN also shows much smoother rise due to its compactness with reduced porosity (inset of **Fig. 3d**).

Regarding the $K$ value, there are several factors which affect the ultimate value such as crystalline structures, orientations and grain sizes. Mateti *et al*, reported that the h-BN sintered at 1800 °C showed $K$ of ~25 W/mK, which enhances with the sintering temperature [5]. With increasing SPS temperature, the grain sizes and density increase with less porosity, resulting in reduced phonon scattering and higher $K$. Recently, it was also shown that h-BN sintered at 1700 °C, resulting in a $K$ ~3.3 W/mK [40]. Effective phonon scattering by pores (high pore density reduces the mean free paths of the phonon) is a factor that reduces $K$. Interestingly, it was found that for twisted h-BN layers, the $K$ becomes even much lower than pristine non-twisted h-BN layers [41,42]. Therefore, the sintering temperature/pressure (which affects the crystallinity and microstructure) is crucial and we note $K$ value commensurate with similar SPS conditions. The $K$ value for SPS h-BN is attributed to the enhanced thermal diffusivity and reduced phonon-scattering, a consequence of the twisted layers and inter-granular fusion resulting in less porosity and enhanced thermal transport [41,43]. In addition, with an increase in density and inter-grain fusion, the surface area among adjacent particles increases along with enhanced cross-linking which significantly reduces the interfacial thermal resistance [44]. Furthermore, for SPS h-BN, from XRD we have seen that the basal planes of the h-BN particles are oriented along the out-of-plane direction of the disk. Since h-BN particles possess highly anisotropic thermal conductivity, thus this would also contribute to enhance the $K$. High $K$ of dense SPS h-BN ceramic offers promises in critical thermal management and heat dissipation for controlled performances.

***Neutron radiation shielding phenomena of SPS h-BN***

As h-BN is a mechanically robust B-rich material, we investigated the performance of SPS h-BN in neutron shielding applications. With the renewed global focus towards nuclear microreactors, nuclear-powered submarines, and the safe transportation of spent nuclear waste, there is a drastic need for a high efficiency neutron shield that is lightweight and mechanically robust [45,46]. Using a monochromatic thermal neutron beam with an incident energy of $E_i$ =16.75 meV ($\lambda$ =2.21 Å) and an incident flux of ~$5\times10^5$ neutrons/cm$^2$/s, we evaluate its shielding performance in comparison to conventional shielding and scattering materials (**Fig. 4a**). The thermal neutron shielding is quantified by the neutron attenuation coefficient $\mu = \ln\frac{I}{I_0}(-t^{-1})$ where $I_0$ and I are the incident and transmitted beam intensities, and $t$ is the material thickness (in cm scale).

The SPS h-BN shows exceptional shielding capabilities with $\mu$ =42.75 cm$^{-1}$ that result in 98.25% neutron shielding for 880 μm material thickness (**Fig. 4b**). This value of $\mu$ is 89% of the ideal theoretical value for stoichiometric h-BN with natural isotopic abundance (80% $^{11}$B and 20% $^{10}$B) indicating a near-perfect density of h-BN particles with few voids or defects within the material [47]. This absorption coefficient is also much higher than the CS h-BN absorption measurements, likely due to void density within the material [48,49]. Additionally, the $\mu$ value shows no thickness dependence as expected for high density materials (**Fig. S10**). It should be noted that besides strong neutron radiation, high temperature is usually present in the reactor. However, reactor applications typically include temperatures from room temperature up to ~300 °C and pressures up to ~11 MPa. Considering h-BN's excellent thermal stability, XRD from room temperature to 500 °C shows no significant structural changes (**Fig. S8**) and thus the temperature dependence of the neutron attenuation coefficient changes will be negligible. Conventionally, to achieve lightweight neutron shielding, h-BN or atomic B have been incorporated in polymer composites showing as high as $\mu$ =34 $cm^{-1}$ [50]. In contrast, the combined mechanically robustness and neutron shielding of pure h-BN ceramics prepared by SPS indicate that pure h-BN serves as lighter and stiffer high-performance materials for neutron attenuation.

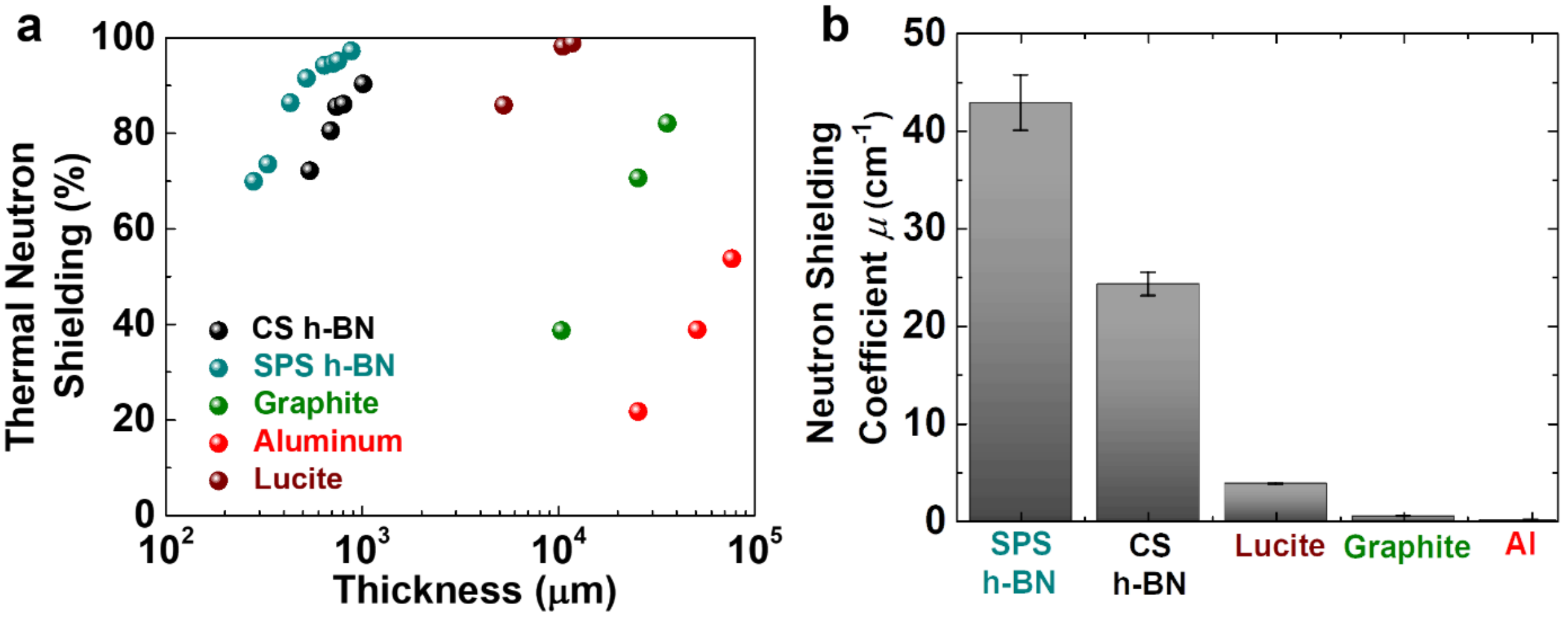

**FIG. 4 Thermal neutron radiation shielding of SPS h-BN**. **(a)** Thermal neutron shielding of spark plasma sintered h-BN (SPS h-BN) ceramic disks and their comparisons with the common neutron absorption and scattering materials. **(b)** Neutron attenuation coefficients, highlighting the outstanding shielding efficiency of SPS h-BN.

The high-density and consequent exceptional functional properties (**Supporting Table S1, S2, and S3**) also raises fundamental question about the structural stability of h-BN (e.g. changes in layer stacking with possible twist, lattice distortion, predominantly $sp^2$ bonded h-BN with some fraction of $sp^3$ bonded c-BN or the metastable super-hard wurtzite *w*-BN due to applied internal or external stimuli [3,51,52]. SPS h-BN shows lattice distortion and moiré-like pattern, however, we did not see any signature of w-BN or c-BN related peaks in the Raman spectroscopy (inset of **Fig.**

**S11**), except the highly-intense $E_{2g}$ h-BN peak throughout the sample, which indicates that the structure-property correlations are associated to pure phase crystalline h-BN. In addition, once the dense h-BN is formed it does not show structural changes and remains highly stable over time, without other phases (**Fig. S12**). To correlate the SPS parameter dependent structural variations with properties, if any, we also sintered samples with varied temperatures and pressure (e.g. 1700 °C/45 MPa and 1600 °C/90 MPa), and corresponding structure-properties are summarized (**Fig. S13** and **Table S4**). In agreement with other reports (**Table S3**), with increasing temperature, the grain sizes and density increase [5,40]. As shown, with increasing SPS temperature, the h-BN shows increase in non-basal plane crystallinity and grain sizes, and produces different twisted layer (**Fig. S13**) [3]. Lower temperature/pressure SPS h-BN samples does not show any significant changes in lattice parameters or cell volume compared to the CS h-BN, and both have almost similar grain sizes (whereas the 1700 °C/90 MPa sintered sample shows much larger grain size with lesser porosity (**Table S4**)). We also measured the mechanical and dielectric properties, also showing the enhancement/modulation with increasing temperature/pressure (**Table S4**). Notably, $E$ increases with increasing temperature, while $H$ increases with pressure, while the combination of both is achieved at 1700ºC and 90 MPa SPS conditions. Twist coming as an additional factor helps in providing more mechanical strength, but might affect in reducing the $K$ [3,41]. All these shows a strong correlation between structure-properties of the resulting h-BN ceramics which can be tuned by the SPS parameters.

We also performed density functional theory (DFT) calculations of distorted h-BN lattice, in order to investigate the change in mechanical and dielectric properties (**Figs. S14** and **S15**). From calculations, the stiffness values are found to be the almost same for both pristine and distorted h-BN, ~44.6 GPa (out-of-plane), similar to previous reports [53]. Our calculations also suggest that the SPS h-BN is close to the ideal monocrystal modelled by DFT. Furthermore, the slight lattice distortions that was experimentally observed do not lead to the high dielectric properties as calculated theoretically. The unique mechanical properties were related to twisting of layers and this structural feature could also be responsible for the anomalous properties that we have observed in the SPS h-BN. We have performed DFT calculation based on twist angle as well, which does not show any significant impact on $E$ and $\kappa$ (**Fig. S16**) [3]. As mentioned earlier, twisted h-BN shows ferroelectricity, which can exhibit spontaneous polarization that can be reoriented by an external electric field, contributing to a higher dielectric response. However, SPS h-BN shows the absence of polarization (which may not nullify its presence at the nanoscale). The high-$\kappa$ than the bulk thus could be due to an increase in ionicity due to a change in local charge distribution as a result of unique microstructures. In addition, the grain sizes and grain boundary can also induce the interfacial polarization. Thus, at present, one can hypothesize that synergistic effects of lattice distortion, layer twisting, non-basal plane crystallinity increases the polarizability (mainly ionic), confinement effects and enhanced interfacial polarization which possibly increases the dielectric constant. Nevertheless, it requires further experiential and computational study which will enable careful optimization for these high-promise bulk ceramics.

## Conclusion

We obtained near-theoretically dense pure phase crystalline bulk h-BN ceramics from micron-scale h-BN powder, by using a high-temperature spark plasma sintering process. We demonstrate SPS h-BN's anomalous properties, beyond intrinsic theoretical predictions, due to the synergistic effects of SPS induced non-basal plane crystallinity, twisting of layers, and inter-grain fusion. These findings are significant for two-dimensional h-BN due to its potential applications spanning manufacturing tools, energy storage devices, critical thermal management, aerospace engineering, and radiation shielding safety for nuclear energy. Our methodology highlights the novel approach in producing high-density bulk materials by twisting of 2D layers and through-thickness interconnections, exhibiting distinct properties amongst ceramic materials.

**Acknowledgements**

This work was sponsored by the Army Research Office and was accomplished under Cooperative Agreement Number W911NF-19-2-0269. The views and conclusions contained in this document are those of the authors and should not be interpreted as representing the official policies, either expressed or implied, of the Army Research Office or the U.S. Government. The U.S. Government is authorized to reproduce and distribute reprints for Government purposes notwithstanding any copyright notation herein. Authors would also like to thank the SPS facility at Texas A & M University, TX, USA. We would also like to acknowledge the SEA facility at Rice University for various characterizations. P. M. A would like to thank Prof. A. Garg for useful discussions

The authors wish to thank the support of the Natural Sciences and Engineering Research Council of Canada, The Vanier Canada Graduate Scholarship, and The Connaught Fund at the University of Toronto. The authors also wish to thank the facility support of the Ontario Centre for the Characterization of Advanced Materials (OCCAM). Use of the MAD beamline at the McMaster Nuclear Reactor is supported by McMaster University and the Canada Foundation for Innovation.

This work was supported by the Department of the Navy, Office of Naval Research, grant number: N00014-22-1-2357; National Science Foundation Graduate Research Fellowship, grant number: 1650114. This work was supported in part by CHIMES, one of seven centers in JUMP 2.0, a Semiconductor Research Corporation (SRC) program sponsored by DARPA.

PGD and CVS acknowledge the Research Alliance of Canada for computational resources. PGD and CVS acknowledges the Natural Sciences and Engineering Research Counsel (NSERC), University of Toronto. PGD acknowledges the Ontario Graduate Scholarship (OGS), and the Vanier Canada Graduate Scholarship for funding.

## AUTHOR DECLARATIONS

### Competing interests

The authors declare that they have no known competing financial interests or personal relationships that could have appeared to influence the work reported in this paper.

## Author contributions

A. B., R. V., and P. M. A. conceptualized the study. A. B., T. S. P., C. L., T. G., B. G., X. Z., A. B. P., J. M., and P. D. synthesized and characterized the materials. J. S. and L. W. M measured dielectric properties. G. A. and Z. T. measured thermal conductivity. P. G. D. and C. S. performed the density fucntioanl theory calculations. P. S., M. H., J. K., B. Y., J. P. C., J. H., Y. Z., and T. F. performed the nanoindentation, micropillar fabrication, mechanical characterizations and neutron absorption studies. A. B. would like to thank Dr. Jianhua Li for his help in high temperature XRD and Mingfei Xu for his help in dielectric measurments. All the authors discussed the results and contributed in manuscript preparation.

## Data Availability Statement

The data that support the findings of this study are available from the corresponding author upon reasonable request.

## Supplementary material

Supplementary material contains HRTEM, XRD, Rietveld refinement, FESEM, Raman spectroscopy, FTIR, thickness dependent neutron attenuation coefficient, property comparison tables, and theoretical calculations.

## References

[1] Z. Dai, L. Liu, Z. Zhang, **Strain Engineering of 2D Materials: Issues and Opportunities at the Interface**, Adv. Mater. 31 (2019) 1805417.

[2] E.Y. Andrei, D.K. Efetov, P. Jarillo-Herrero, A.H. MacDonald, K.F. Mak, T. Senthil, E. Tutuc, A. Yazdani, A.F. Young, **The marvels of moiré materials**, Nat. Rev. Mater. 6 (2021) 201–206.

[3] Y. Wu, Y. Zhang, X. Wang, W. Hu, S. Zhao, T. Officer, K. Luo, K. Tong, C. Du, L. Zhang, B. Li, Z. Zhuge, Z. Liang, M. Ma, A. Nie, D. Yu, J. He, Z. Liu, B. Xu, Y. Wang, Z. Zhao, Y. Tian, **Twisted-layer boron nitride ceramic with high deformability and strength**, Nature 626 (2024) 779–784.

[4] W. Ouyang, H. Qin, M. Urbakh, O. Hod, **Controllable Thermal Conductivity in Twisted Homogeneous Interfaces of Graphene and Hexagonal Boron Nitride**, Nano Lett. 20 (2020) 7513–7518.

[5] S. Mateti, K. Yang, X. Liu, S. Huang, J. Wang, L. Hua, H. Li, P. Hodgson, M. Zhou, J. He, Y. Chen, **Bulk Hexagonal Boron Nitride with a Quasi-Isotropic Thermal Conductivity**, Adv. Func. Mater. 28 (2018) 1707556.

[6] X. Duan, Z. Yang, L. Chen, Z. Tian, D. Cai, Y. Wang, D. Jia, Y. Zhou, **Review on the properties of hexagonal boron nitride matrix composite ceramics**, J. Eur. Ceram. Soc. 36 (2016) 3725–3737.

[7] B. Niu, D. Jia, D. Cai, Z. Yang, X. Duan, W. Duan, Q. Li, B. Qiu, P. He, Y. Zhou, **Grain-orientation dependence of the anisotropic thermal shock performance of hexagonal boron nitride ceramics**, Scr. Mater. 178 (2020) 402–407.

[8] A. E. Naclerio, P. R. Kidambi, **A Review of Scalable Hexagonal Boron Nitride (h-BN) Synthesis for Present and Future Applications**, Adv. Mater. 35 (2022) 2207374.

[9] A. Laturia, M. Van de Put, W. Vandenberghe, **Dielectric properties of hexagonal boron nitride and transition metal dichalcogenides: from monolayer to bulk**, npj 2D Materials and Applications 2018, 2 (2018) 6.

[10] J.T. Cahill, W.L. Du Frane, C.K. Sio, G.C.S. King, J.C. Soderlind, R. Lu, M.A. Worsley, J.D. Kuntz, **Transformation of boron nitride from cubic to hexagonal under 1-atm helium,** Diam. Relat. Mater. 109 (2020) 108078.

[11] H. Yang, H. Fang, H. Yu, Y. Chen, L. Wang, W. Jiang, Y. Wu, J. Li, **Low temperature self-densification of high strength bulk hexagonal boron nitride**, Nat. Commun. 2019, 10 (2019) 854.

[12] J. Eichler, C. Lesniak, **Boron nitride (BN) and BN composites for high-temperature applications**, J. Eur. Ceram. Soc. 28 (2008) 1105–1109.

[13] T.B. Wang, C.C. Jin, J. Yang, C.F. Hu, T. Qiu, **Physical and mechanical properties of hexagonal boron nitride ceramic fabricated by pressureless sintering without additive**, Adv. Appl. Ceram. 114 (2015) 273–276.

[14] J. Zhang, R. Tu, T. Goto, **Preparation of Ni-precipitated hBN powder by rotary chemical vapor deposition and its consolidation by spark plasma sintering**, J. Alloys Compd. 502 (2010) 371–375.
[15] F.R. Zhai, M. Lu, K. Shan, Z.Z. Yi, Z.P. Xie, **Spark Plasma Sintering and Characterization of Mixed h-BN Powders with Different Grain Sizes**, Solid State Phenom. 281 (2018) 414–419.
[16] M. Ehsani, M. Zakeri, M. Razavi, **The effect of boron oxide on the physical and mechanical properties of nanostructured boron nitride by spark plasma sintering**, J. Alloys Compd. 780 (2019) 570–573.
[17] O.A.M. Elkady, A. Abu-Oqail, E.M.M. Ewais, M. El-Sheikh, **Physico-mechanical and tribological properties of Cu/h-BN nanocomposites synthesized by PM route**, J. Alloys Compd. 625 (2015) 309–317.
[18] Z.A. Munir, U. Anselmi-Tamburini, M. Ohyanagi, **The effect of electric field and pressure on the synthesis and consolidation of materials: A review of the spark plasma sintering method**, J. Mater. Sci. 41 (2006) 763–777.
[19] B.H. Toby, R.B. Von Dreele, **GSAS-II: The Genesis of a Modern Open-Source All-Purpose Crystallography Software Package**, J. Appl. Crystallogr. 46 (2013) 544–549.
[20] G. Kresse, J. Hafner, **Ab initio molecular dynamics for liquid metals**, Phys. Rev. B 47 (1993) 558–561.
[21] S. Saha, A. Rice, A. Ghosh, S. M. N. Hasan, W. You, T. Ma, A. Hunter, L. J. Bissell, R. Bedford, M. Crawford, S. Arafin, **Comprehensive characterization and analysis of hexagonal boron nitride on sapphire**, AIP Advances 11 (2021) 055008.
[22] C. R. Woods, P. Ares, H. Nevison-Andrews, M. J. Holwil, R. Fabregas, F. Guinea, A. K. Geim, K. S. Novoselov, N. R. Walet, L. Fumagalli, **Charge-polarized interfacial superlattices in marginally twisted hexagonal boron nitride**, Nat. Commun 12 (2021) 347.
[23] K. Yasuda, X. Wang, K. Watanabe, T. Taniguchi, P. Jarillo-herrero, **Stacking-engineered ferroelectricity in bilayer boron nitride**, Science 372 (2021) 1458.
[24] J. K. Hite, Z. R. Robinson, Jr. C. R. Eddy, B. N. Feigelson, **Electron Backscatter Diffraction Study of Hexagonal Boron Nitride Growth on Cu Single-Crystal Substrates**, ACS Appl. Mater. Interfaces 7, 28 (2015) 15200–15205.
[25] O. Guillon, J. Gonzalez-Julian, B. Dargatz, T. Kessel, G. Schierning, J. Räthel, M. Mathias Herrmann, **Field-Assisted Sintering Technology/Spark Plasma Sintering: Mechanisms, Materials, and Technology Developments**, Adv. Eng. Mater. 16 (2014) 830.
[26] A. Loganathan, A. Sharma, C. Rudolf, C. Zhang, P. Nautiyal, S. Suwas, B. Boesl, A. Agarwal, **In-situ deformation mechanism and orientation effects in sintered 2D boron nitride nanosheets**, Mater. Sci. Eng. A 708 (2017) 440–450.
[27] X. Duan, Z. Yang, L. Chen, Z. Tian, D. Cai, Y. Wang, D. Jia, Y. Zhou, **Review on the properties of hexagonal boron nitride matrix composite ceramics**, J. Eur. Ceram. Soc. 36 (2016) 3725–3737.

[28] X. Duan, D. Jia, Z. Wang, D. Cai, Z. Tian, Z. Yang, P. He, S. Wang, Y. Zhou, **Influence of hot-press sintering parameters on microstructures and mechanical properties of h-BN ceramics**, J. Alloys Compd. 684 (2016) 474–480.

[29] J. Zou, G.-J. Zhang, Z.-J. Shen, J. Binner, **Ultra-low temperature reactive spark plasma sintering of $ZrB_2$-hBN ceramics**, J. Eur. Ceram. Soc. 36 (2016) 3637–3645.

[30] Y. Yang, Z. Song, G. Lu, Q. Zhang, B. Zhang, B. Ni, C. Wang, X. Li, L. Gu, X. Xie, H. Gao, J. Lou, **Intrinsic toughening and stable crack propagation in hexagonal boron nitride**, Nature 594 (2021) 57–61.

[31] A.Z. Juri, A.K. Basak, L. Yin, **Failure mechanisms in in-situ SEM micropillar compressions of pre-crystallized and crystallized zirconia-containing lithium silicate glass-ceramics**, Ceram. Int. 49 (2023) 27165–27175.

[32] J. D. Giallonardon, U. Erb, K. T. Aust, G. Palumbo, **The influence of grain size and texture on the Young's modulus of nanocrystalline nickel and nickel–iron alloys**, Phil. Mag. 91 (2011) 4594-4605.

[33] C.R. Dean, A.F. Young, I. Meric, C. Lee, L. Wang, S. Sorgenfrei, K. Watanabe, T. Taniguchi, P. Kim, K.L. Shepard, J. Hone, **Boron nitride substrates for high-quality graphene electronics**, Nat. Nanotechnol. 5 (2010) 722–726.

[34] T. Knobloch, Y.Y. Illarionov, F. Ducry, C. Schleich, S. Wachter, K. Watanabe, T. Taniguchi, T. Mueller, M. Waltl, M. Lanza, M.I. Vexler, M. Luisier, T. Grasser, **The performance limits of hexagonal boron nitride as an insulator for scaled CMOS devices based on two-dimensional materials**, Nat. Electron. 4 (2021) 98–108.

[35] A.A. Bokov, **Dielectric relaxation in relaxor ferroelectrics**, Z.-G. Ye, J. Adv. Dielectr. 02 (2012) 1241010.

[36] M. J Meziani, W. Song, P. Wang, F. Lu, Z. Hou, A. Anderson, H. Maimaiti, Y. Sun, **Boron Nitride Nanomaterials for Thermal Management Applications**, ChemPhysChem 16 (2015) 1339-1346.

[37] L. Dusza, **Combined solution of the simultaneous heat loss and finite pulse corrections with the laser flash method**, High Temp.-High Press. 27/28 (1995) 467–473.

[38] Q. Zheng, M. Hao, R. Miao, J. Schaadt, C. Dames, **Advances in thermal conductivity for energy applications: a review**, Prog. Energy 3 (2021) 012002.

[39] P. Jiang, X. Qian, R. Yang, L. Lindsay, **Anisotropic thermal transport in bulk hexagonal boron nitride**, Phys. Rev. Mater. 2 (2018) 064005.

[40] Z. Yılmaz, N. Ay, **The investigation of synthesis and textured properties of in situ formed hBN with spark plasma sintering**, Mater. Chem. Phys. 316 (2024) 129043.

[41] G.R. Jaffe, K.J. Smith, K. Watanabe, T. Taniguchi, M.G. Lagally, M.A. Eriksson, V.W. Brar, **Thickness-Dependent Cross-Plane Thermal Conductivity Measurements of Exfoliated Hexagonal Boron Nitride**, ACS Appl. Mater. Interfaces 15 (2023) 12545–12550.

[42] A. Biswas, R. Xu, G. A. Alvarez, J. Zhang, J. Christiansen-Salameh, A. B Puthirath, K. Burns, J. A Hachtel, T. Li, S. A. Iyengar, T. Gray, C. Li, X. Zhang, H. Kannan, J. Elkins, T. S Pieshkov, R. Vajtai, A. G. Birdwell, M. R Neupane, E. J Garratt, T. Ivanov, B. B Pate, Y.

Zhao, H. Zhu, Z. Tian, A. Rubio, P. M Ajayan, **Non-Linear Optics at Twist Interfaces in h-BN/SiC Heterostructures**, Adv. Mater. 35 (2023) 2304624.

[43] J.-X. Xue, J.-X. Liu, B.-H. Xie, G.-J. Zhang, **Pressure-induced preferential grain growth, texture development and anisotropic properties of hot-pressed hexagonal boron nitride ceramics**, Scr. Mater. 65 (2011) 966–969.

[44] R. Mo, Z. Liu, W. Guo, X. Wu, Q. Xu, Y. Min, J. Fan, J. Yu, **Interfacial crosslinking for highly thermally conductive and mechanically strong boron nitride/aramid nanofiber composite film**, Compos. Commun. 28 (2021) 100962.

[45] Z. Qi, Z.Yang, J. Li, Y. Guo, G. Yang, Y. Yu, J. Zhang, **The Advancement of Neutron-Shielding Materials for the Transportation and Storage of Spent Nuclear Fuel**, Materials 15 (2022) 3255.

[46] S. Hasan, I. Ahmad, **Progress in Hexagonal Boron Nitride (h-BN)-Based Solid-State Neutron Detector**, Electron. Mater. 3(3) (2022) 235-251.

[47] A. Mballo, A. Ahaitouf, S. Sundaram, A. Srivastava, V. Ottapilakkal, R. Gujrati, P. Vuong, S. Karrakchou, M. Kumar, X. Li, Y. Halfaya, S. Gautier, P.L. Voss, J.P. Salvestrini, A. Ougazzaden, **Natural Boron and $^{10}$B-Enriched Hexagonal Boron Nitride for High-Sensitivity Self-Biased Metal–Semiconductor–Metal Neutron Detectors**, ACS Omega 7 (2022) 804–809.

[48] K. Ahmed, R. Dahal, A. Weltz, J. J. -Q. Lu, Y. Danon, I. B. Bhat, **Solid-state neutron detectors based on thickness scalable hexagonal boron nitride**, Appl. Phys. Lett. 110 (2017) 023503.

[49] J. Li, R. Dahal, S. Majety, J.Y. Lin, H.X. Jiang, **Hexagonal boron nitride epitaxial layers as neutron detector materials**, Nucl. Instrum. Methods Phys. Res. Sect. Accel. Spectrometers Detect. Assoc. Equip. 654 (2011) 417–420.

[50] Y. Shang, G. Yang, F. Su, Y. Feng, Y. Ji, D. Liu, R. Yin, C. Liu, C. Shen, **Multilayer polyethylene/ hexagonal boron nitride composites showing high neutron shielding efficiency and thermal conductivity**, Compos. Commun. 19 (2020) 147–153.

[51] C. Chen, D. Yin, T. Kato, T. Taniguchi, K. Watanabe, X. Ma, H. Ye, Y. Ikuhara, **Stabilizing the metastable superhard material wurtzite boron nitride by three-dimensional networks of planar defects**, Proc. Natl. Acad. Sci. 116 (2019) 11181–11186.

[52] Y. Meng, H. Mao, P.J. Eng, T.P. Trainor, M. Newville, M.Y. Hu, C. Kao, J. Shu, D. Hausermann, R.J. Hemley, **The formation of $sp^3$ bonding in compressed BN**, Nat. Mater. 3 (2004) 111–114.

[53] C. Cazorla, T. Gould, **Polymorphism of bulk boron nitride**, Sci. Adv. 5.1 (2019) eaau5832.

## Supplementary material

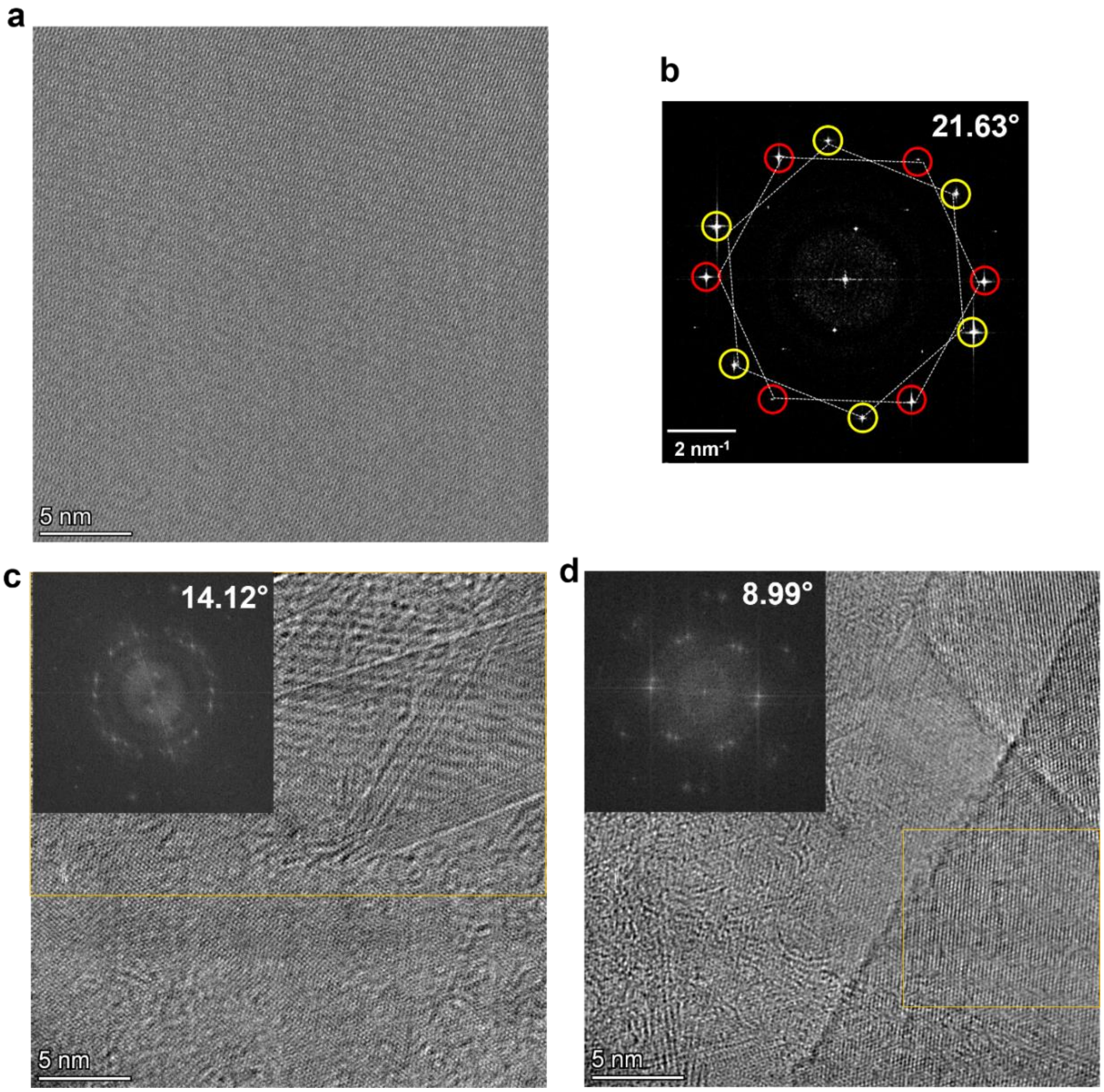

**Fig. S1 Moiré patterns of SPS h-BN. a)** Large-scale HRTEM shows the Moiré structure. **b)** The FFT pattern shows two diffractions spots (marked as red and yellow circles enclosed with dashed hexagons) with a rotation angle between them of ~21.63°. **c), d)** Several other regions also show the Moiré structures (yellow regions) with two diffractions spots, having a rotation angle between them of ~14.12° and ~8.99°.

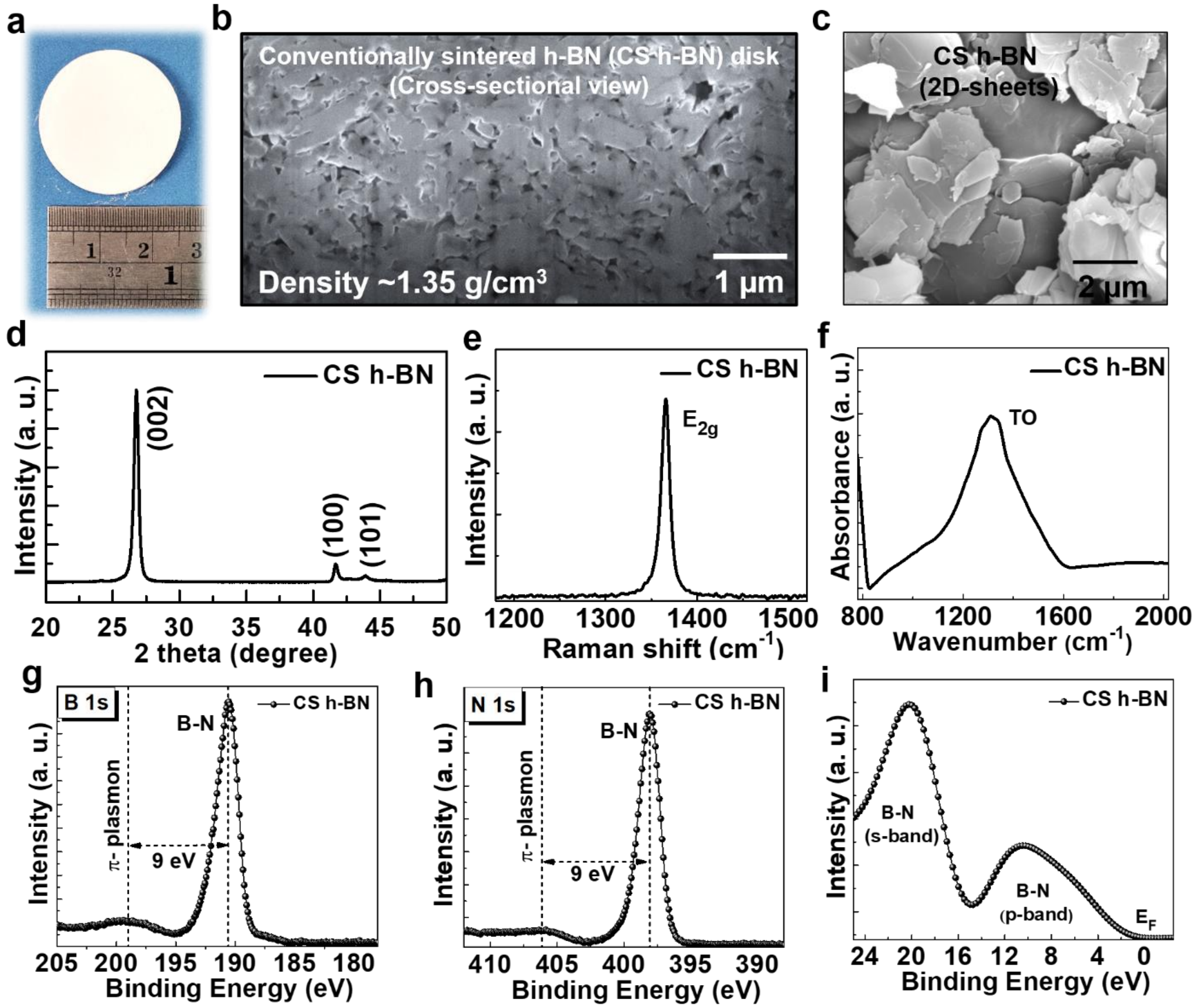

**Fig. S2 Structural characterizations of conventionally sintered h-BN (CS h-BN). a)** Sintered one-inch diameter disk. **b)** Cross-sectional FESEM of the disk. **c)** FESEM show sheet-like features (size ~1-3 μm). **d)** XRD patterns of the disk. **e), f),** Raman spectroscopy shows the in-plane $E_{2g}$ peak and FTIR spectra show the transverse optical (TO) mode peak correspond to h-BN. **g), h)** Core-level B 1s and N 1s XPS scans. **i)** valence band spectra (VBS) of conventionally sintered h-BN.

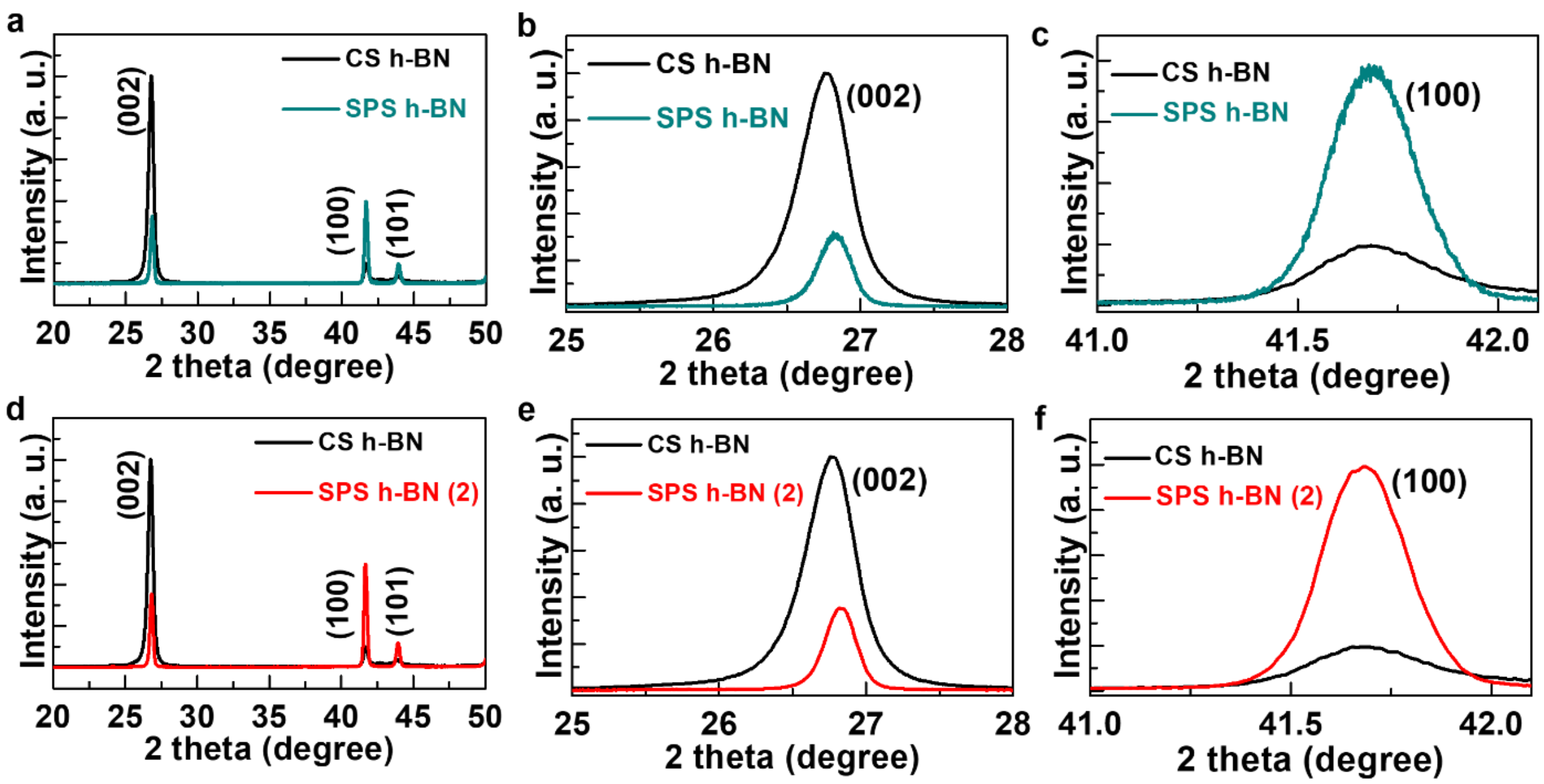

**Fig. S3 XRD of SPS h-BN. a), b), c)** For SPS h-BN, the relative intensity of the non-basal plane (hkl) peak intensity is increased. This indicates a change in preferred grain orientation, while the shift in peak position indicates reduced *d*-spacing along the c-axis. **d), e), f)** XRD reproducibility of SPS h-BN samples. "2" corresponds to another sintered SPS h-BN disk.

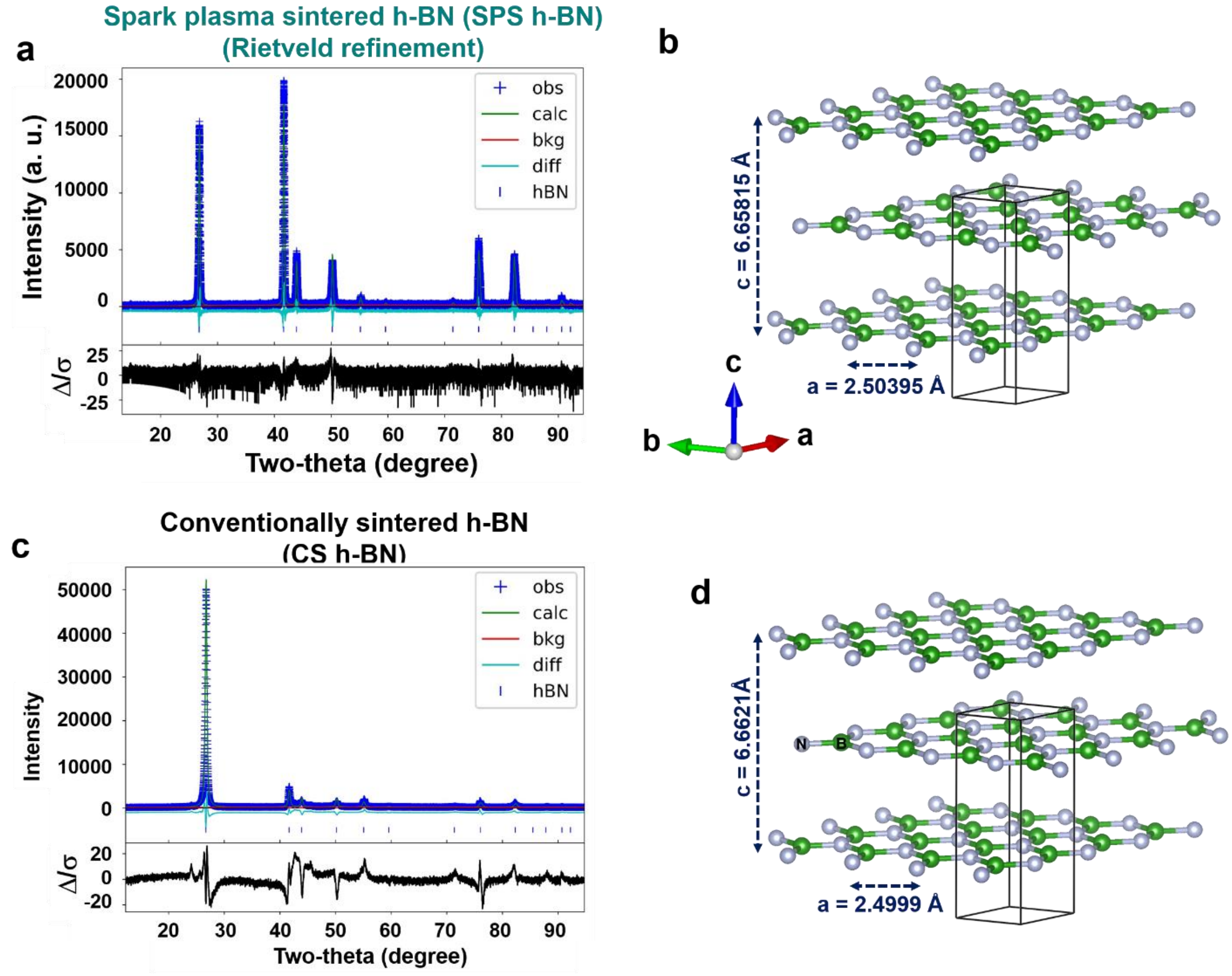

**Fig. S4 Rietveld refinement of SPS h-BN XRD pattern. a)** For SPS h-BN the refined lattice parameters are found to be as follows: $a$ = 2.50395(2) Å, $c$ = 6.65815(7) Å. **b)** The unit cell lattice structures SPS h-BN. **c), d)** For pristine h-BN the refined lattice parameters are $a$ = 2.4999(2) Å, $c$ = 6.6621(2) Å.

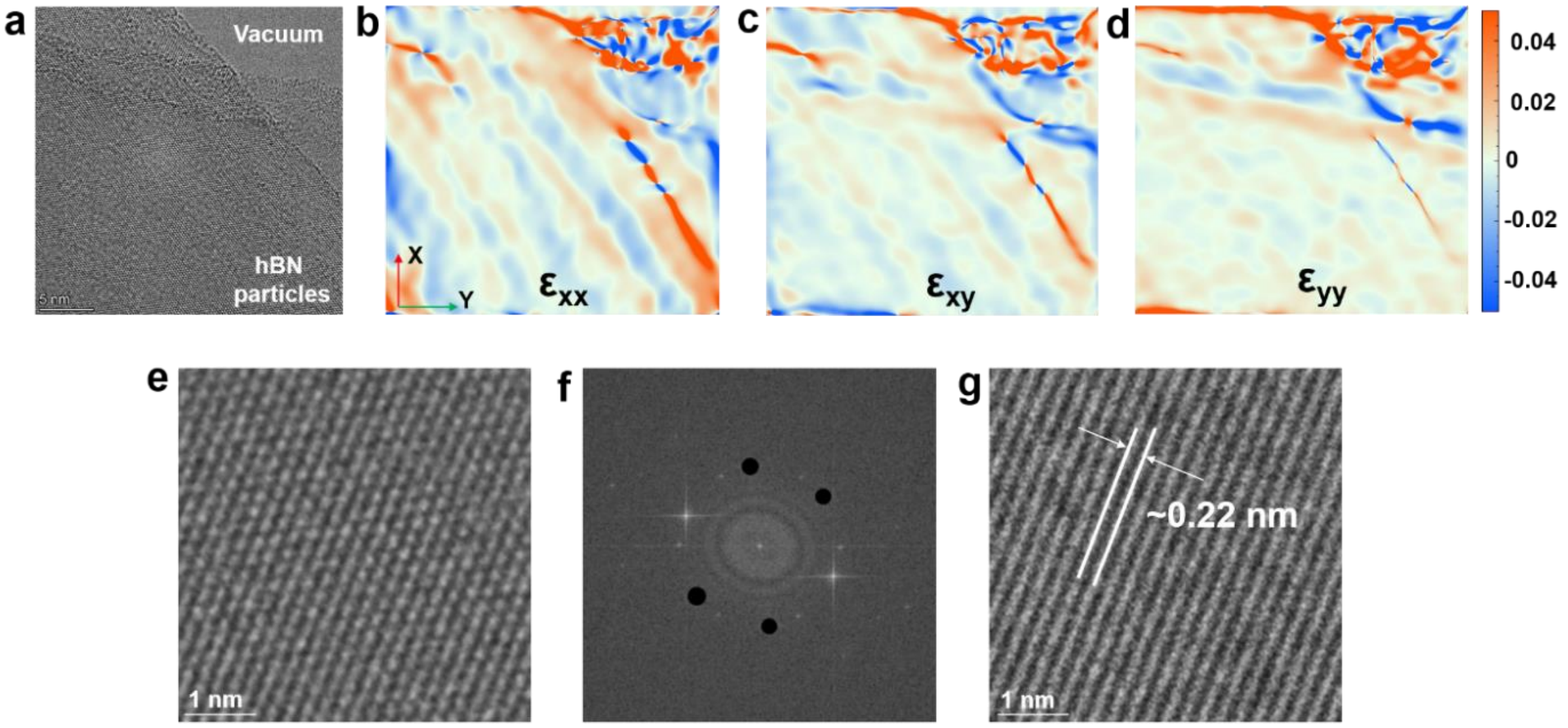

**Fig. S5 HRTEM of SPS h-BN.** TEM analysis of lattice expansion in h-BN particle. **a)** HRTEM image of two h-BN particles overlapped on each other; **b)** strain map along the x-direction; **c)** along the xy-direction; and **d)** along the y-direction. **e)** Atomic resolution TEM of the h-BN particle; **f)** FFT image of the e) with a circular mask covering all but for one set of planes; **g)** atomic resolution TEM image after decomposition with measured d-spacing along the $[10\bar{1}0]$.

To investigate the strain, we conducted geometrical phase analysis which was originally developed by Martin Hytch. The method allows us to see the deformation at the nanoscale and derive the strain field from it by comparing the phase of the planes from the atomic resolution TEM image. We used the Strain ++ software developed be J.J.P. Peters (https://jjppeters.github.io/Strainpp/) for this type of analysis. HRTEM shows the image of the h-BN particle (**Fig. 5a**). Furthermore, strain mapping (**Figs. 5b-d**) along the *x*-direction ($\varepsilon_{xx}$), diagonal between *x* and *y* axes ($\varepsilon_{xy}$), and *y*-direction ($\varepsilon_{yy}$) respectively, shows the ripples-like features which confirms the presence of strain (<2%) in the sample. The region in the upper right corner over the vacuum shows the artifact strain after calculations since it was measured with reference to the h-BN particle.

We measured the *d*-spacing between the planes $[10\bar{1}0]$ (**Fig. 5g**). For simplifications, measurements, we decomposed the (**Fig. 5e**) image by masking the FFT spots shown on (**Fig. 5f**). As a result, we see only the set of planes, corresponding to $[10\bar{1}0]$ direction. Our calculated *d*-spacing ~0.22 nm is larger than the standard h-BN $[10\bar{1}0]$ *d*-spacing of ~0.2165 (PDF Card No. 00-034-0421) which is the evidence of in-plane lattice expansion in h-BN.

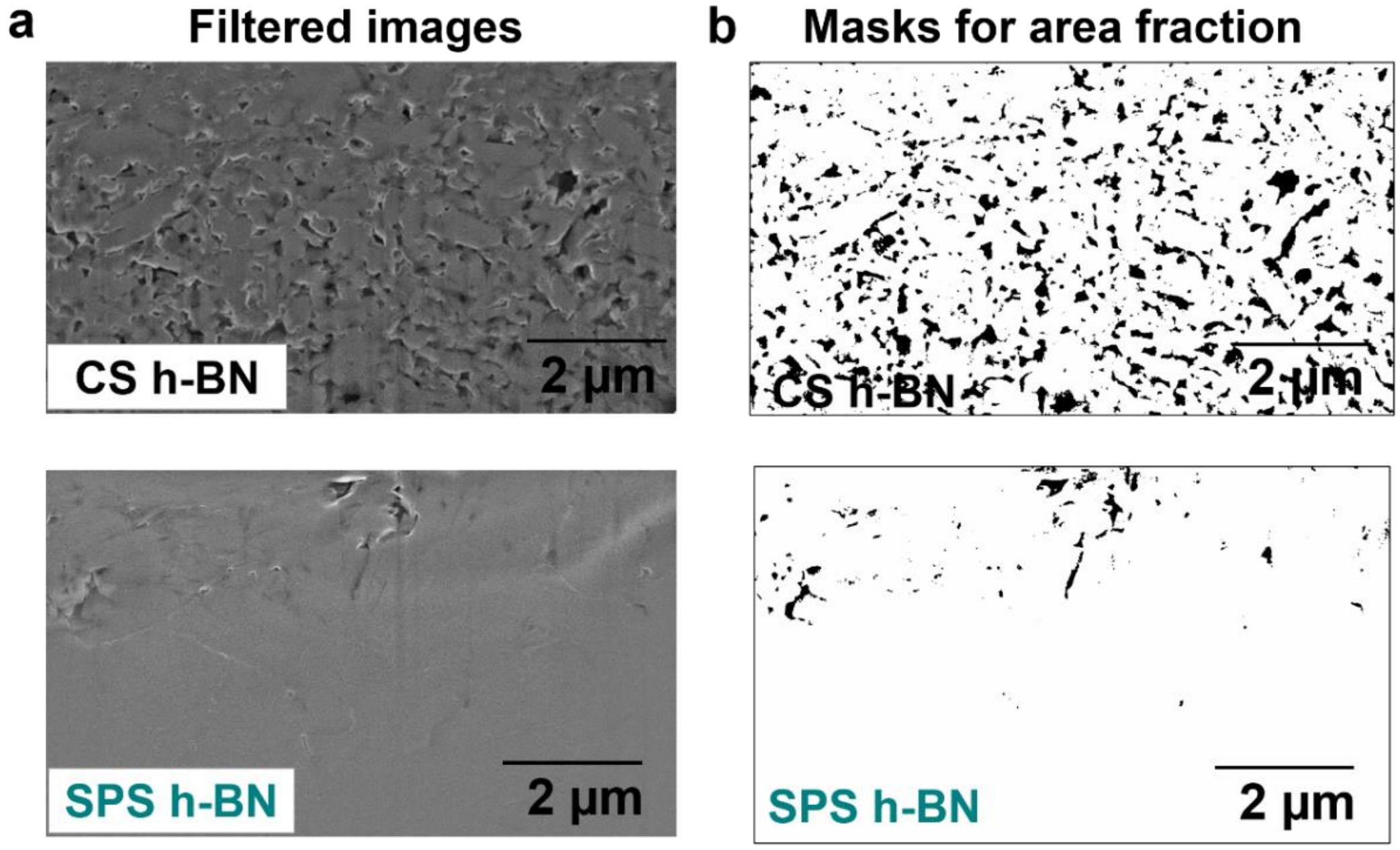

**Fig. S6 Porosity of h-BN disks from cross-sectional FESEM. a), b)** We applied intensity threshold masks to calculate the pore's density from FESEM images. Images were filtered to remove the intensity gradient due to charging of the surface by dividing the original image on the duplicated one with gaussian filter. Next, after adjusting the threshold values of each images they were binarized before conducting measurements of the respective areas. We obtained pore's density of ~1.1% for SPS h-BN, (whereas its ~13.5% for CS h-BN) over the total cross-sectional image area.

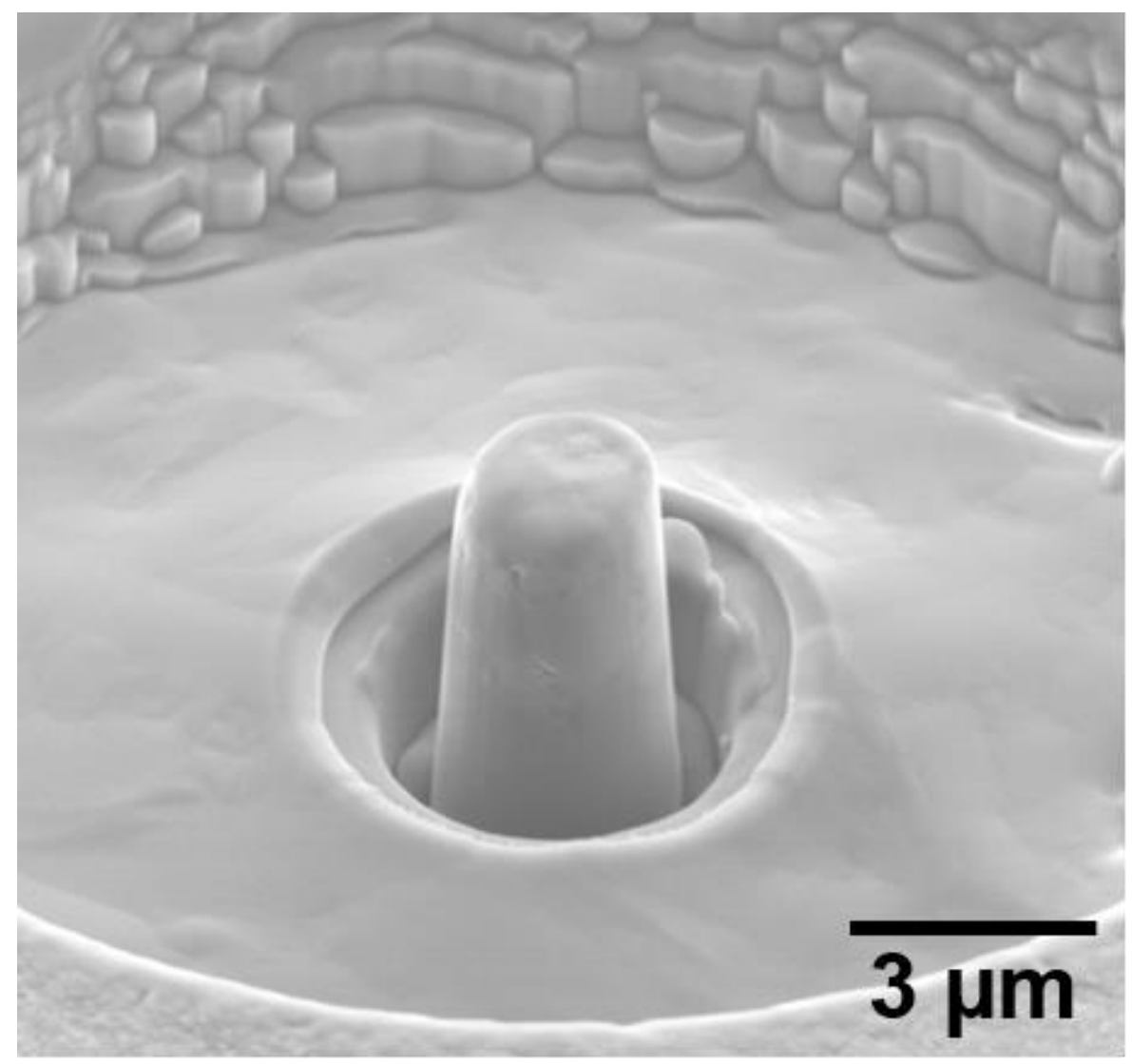

**Fig. S7 Micro-pillars made from SPS h-BN.** As-cut micro-pillar of SPS h-BN made from the SPS h-BN disk by using the focused-ion beam milling process.

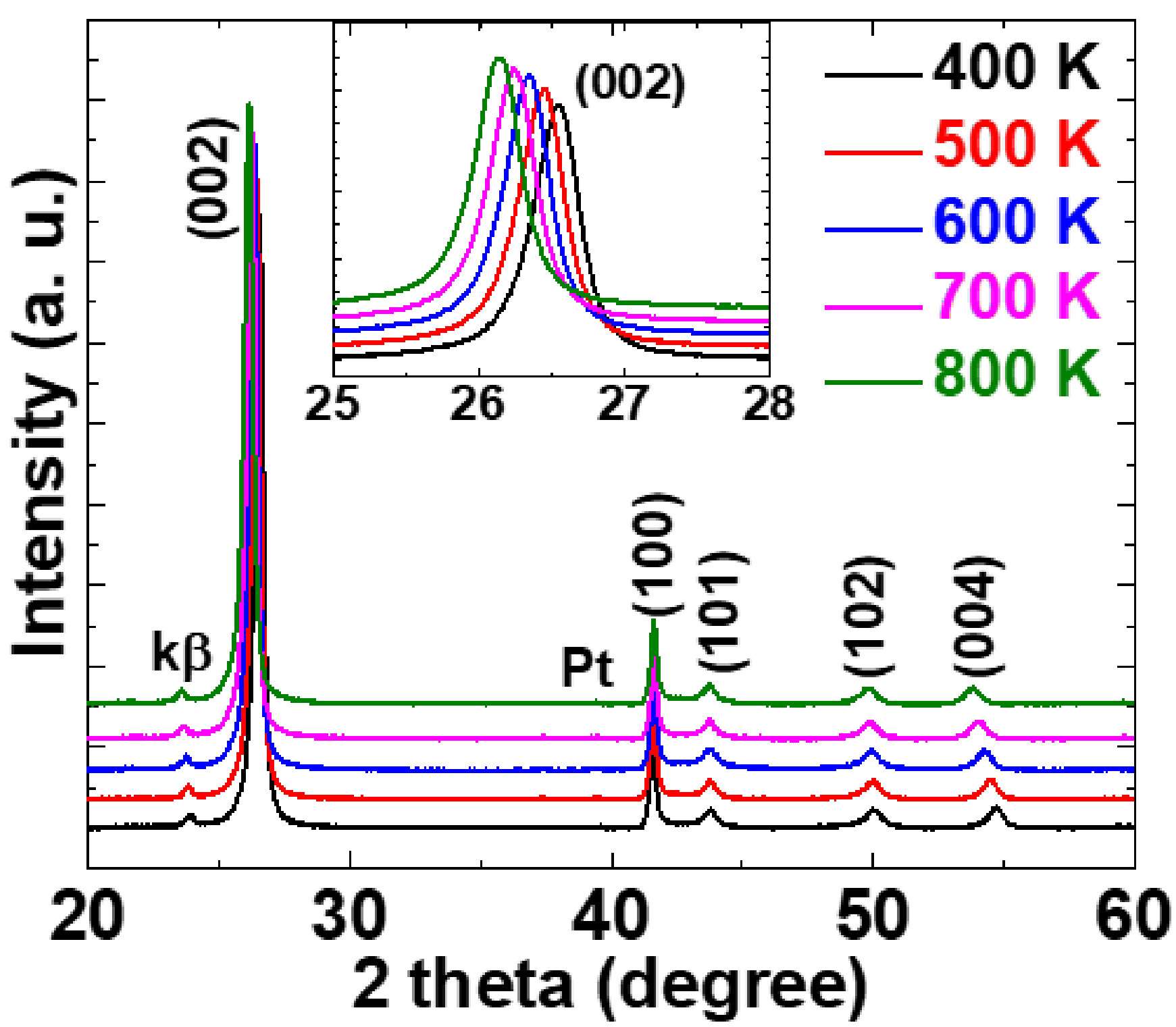

**Fig. S8 Temperature dependent XRD of SPS h-BN.** Temperature dependent XRD within 400-800 K shows no significant changes, except the thermal expansion driven shift in 2θ-values (inset). Due to the instrumental limitation, HT-XRDs were done on powders, scratched from the disk and grounded for 10 min in a mortal pastel to make fine grains. Sample holder Pt peaks are also seen.

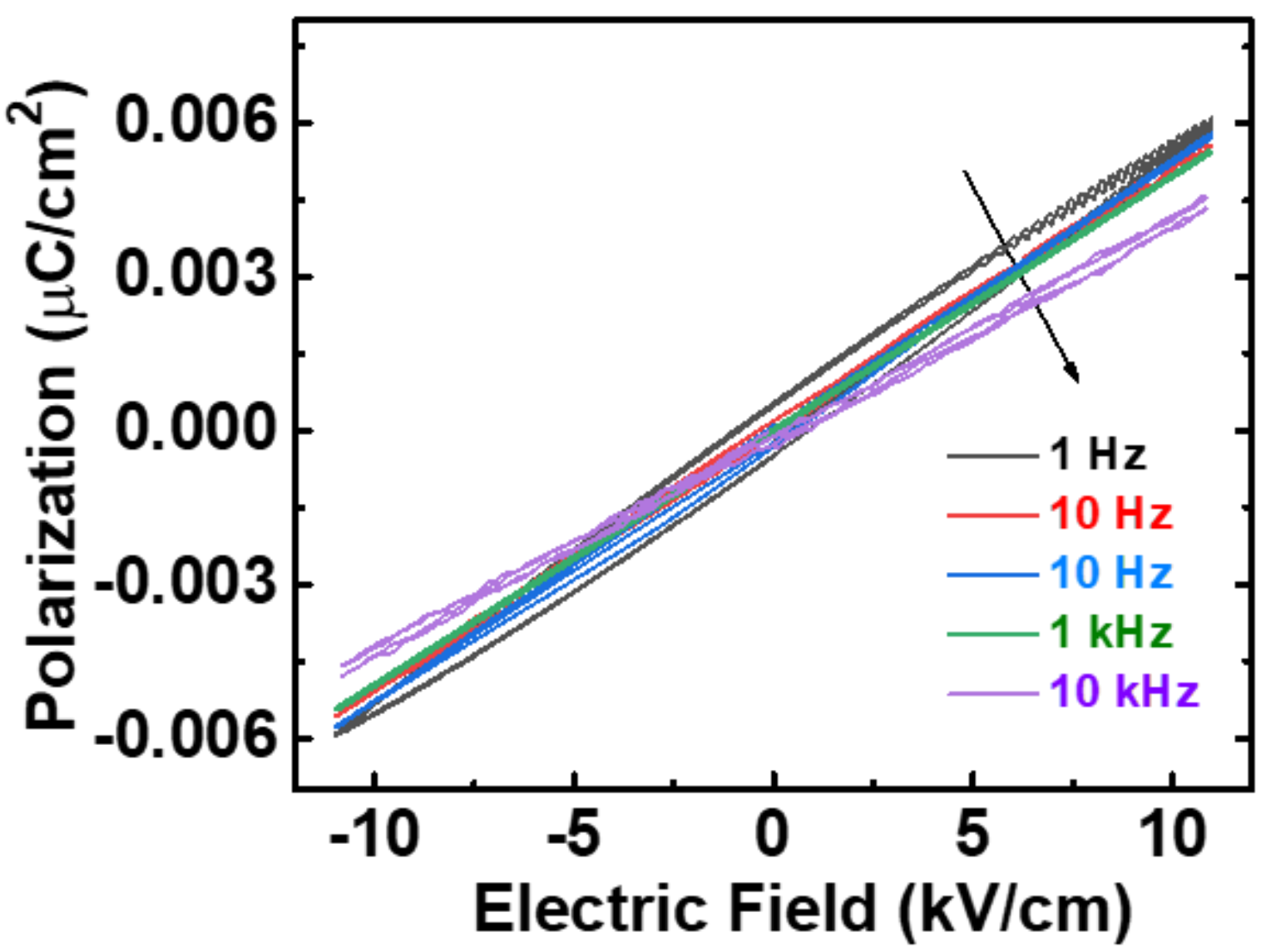

**Fig. S9 Investigation of ferroelectricity in SPS h-BN**. Absence of hysteresis loop in polarization vs electric field within the applied field range.

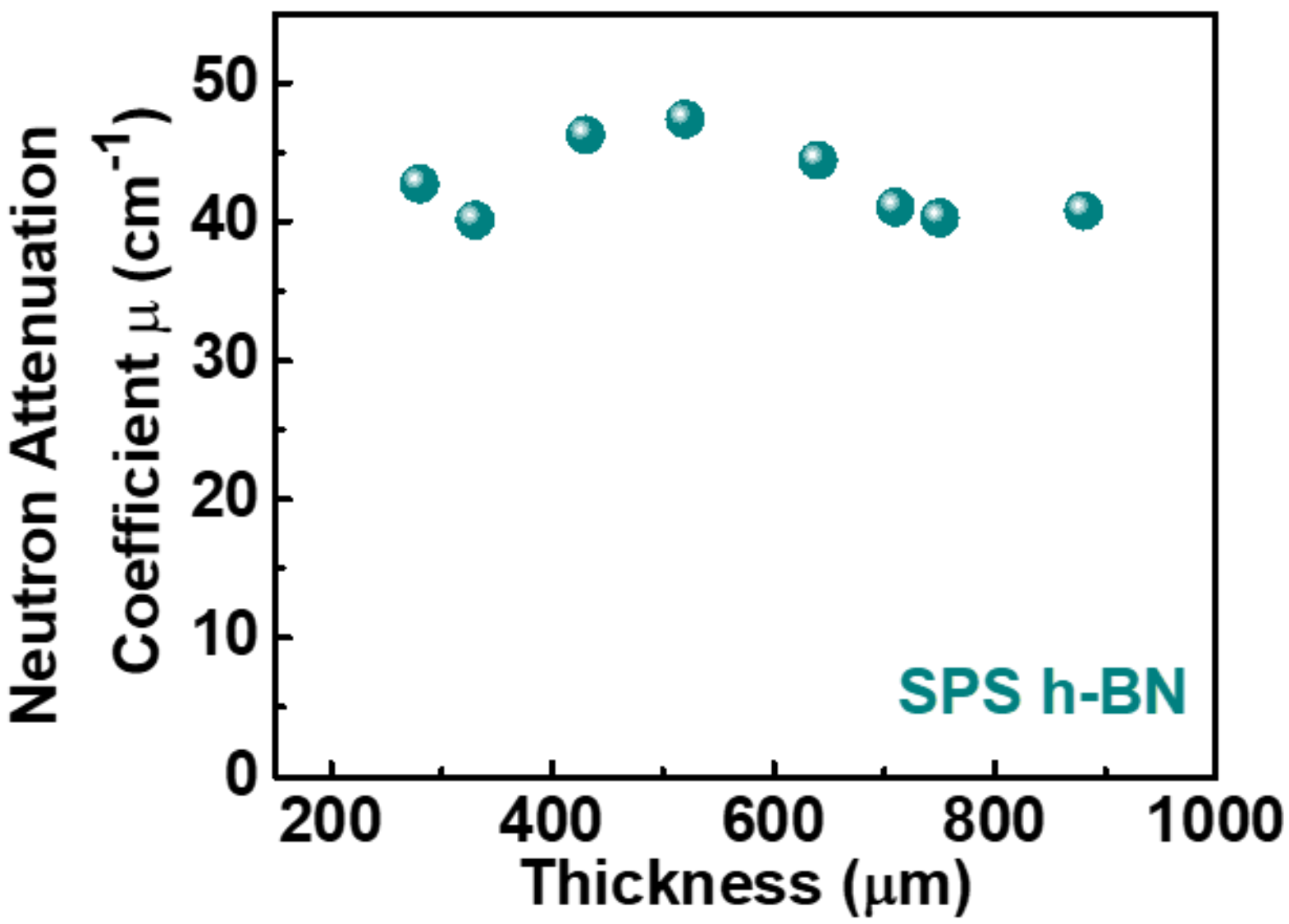

**Fig. S10 Thickness dependent neutron attenuation coefficient ($\mu$) for SPS h-BN.** Thickness dependent neutron attenuation coefficient for SPS h-BN exhibits minimal variability, suggesting that there is no thickness dependency.

**Table S1: Properties of spark plasma sintered (SPS) bulk h-BN.**

| Properties (at room temperature) | SPS h-BN |
| --- | --- |
| **Density ($\rho$)** | $2.05 \pm 0.04$ g/cm$^3$ |
| **Static dielectric constant ($\kappa$)** | 10.78 ($\kappa_{\parallel}$), 9.01 ($\kappa_{\perp}$) at 1 kHz |
| **Thermal conductivity ($K_{\perp}$)** | $18.72 \pm 1.6$ W/mK |
| **Young's modulus ($E$)** | $69.7 \pm 3$ GPa |
| **Hardness ($H$)** | $283.3 \pm 9$ MPa |
| **Yield strength** ($\sigma_y$) | $245.1 \pm 39$ MPa |
| **Failure strain** ($\varepsilon$) | 11.5% |
| **Neutron shielding efficiency** | 98.25 % (for 880 um thickness) |
| **Neutron attenuation coefficient ($\mu$)** | 42.75 cm$^{-1}$ |

**Table S2: Comparison of properties of SPS h-BN with single crystal h-BN.**

| **Sample** | **Lattice spacing (Å)** | **Density (g/cm$^3$)** | **Cross-plane thermal conductivity (W/mK)** | **Young's modulus (GPa)** | **Hardness (MPa)** | **Failure strain (%)** | **Dielectric constant (at 1 kHz)** | **Neutron attenuation coefficient (cm$^{-1}$)** |
|---|---|---|---|---|---|---|---|---|
| SPS h-BN (Bulk) | *a* = 2.504, *c* =6.658 | 2.05 | 18.24 | 69.7 | 283.3 | 11.5 % | $\kappa_{\parallel}$ ~10.78, $\kappa_{\perp}$ ~9.01 | 42.75 |
| Single crystal h-BN (Mono/ few layers) | *a* =2.50, *c* = 6.66 | 2.10 (**Ref [10]**) | 8.1 (**Ref [41]**) | 865 (**Ref [S1]**) | 70.5 GPa (**Ref [S1]**) | 17% (**Ref [S1]**) | $\kappa_{\parallel}$ ~6.82 $\kappa_{\perp}$ ~3.29 (**Ref [9]**) | NA Sample handling limitations of few-layer h-BN |

## Table S3: Comparison of structure and properties of dense SPS h-BN samples.

| Pre-cursor | SPS condition | Density (g/cm$^3$) | Micro-structures | Thermal Conduc-tivity (W/mK) | Young's Modulus (GPa) | Hardness (MPa) | Failure strain (%) | Dielectric constant (at 1kHz) | Neutron attenua-tion coefficient (μ) cm$^{-1}$ | References |
|---|---|---|---|---|---|---|---|---|---|---|
| **h-BN powder** | **1700 °C /90 MPa, 60 min** | **2.05** | **Moiré pattern, non-basal plane crystallinity, lattice distortion** | **18.24** | **69.7** | **283.3** | **11.5%** | **$\kappa_{\parallel}$ ~10.78, $\kappa_{\perp}$ ~9.01** | **42.75** | **Our results** |
| o-BN | 1800 °C /5 MPa, 5-10 min | 2.08 | Out-of-plane compression. /twisted layers | NA | 27.9 | 619 | 14% | NA | NA | **Ref [3]** |
| Exfoliated BNNS | 1800 °C /50 MPa, 5 min | 1.53 | Increase in crystalline size with temp | 25 | NA | 307 | NA | NA | NA | **Ref [5]** |
| 0.5 µm BN + (Mg–Al–Si–O) | 1800 °C /30 MPa, 60 min | 2.3 | Porosity <0.1% | NA | 50.7 | NA | NA | NA | NA | **Ref [7]** |
| h-BN + c-BN powder | 1700 °C /80 MPa, 5 min | 2.1 | c-BN transforms to BN onions and fill voids | NA | 20 (50% c-BN) | 90 | NA | NA | NA | **Ref [11]** |
| h-BN + Ni | 2273 °C /100 MPa, 10 min | ~ 2.03 (with 1.87 mass % Ni) | Ni distributed uniformly on h-BN surface | NA | NA | 300 | NA | NA | NA | **Ref [14]** |
| h-BN powder of different grains | 1600 °C /40 MPa, 5 min | 1.82 (90% sub µm powder) | large grain size can form a higher directional arrangement | NA | 80 | NA | NA | NA | NA | **Ref [15]** |
| h-BN + $B_2O_3$ | 1600 °C /30 MPa, 10 min | 2.1 (3 wt% $B_2O_3$) | Pore size and porosity reduced | NA | NA | 1.67 GPa | NA | 6.2 (at 1 MHz) | NA | **Ref [16]** |
| $H_3BO_3$ + $CO(NH_2)_2$ | 1700 °C /30 MPa, 15 min | 2.09 | Nano and um size grains | 3.3 | NA | NA | NA | NA | NA | **Ref [40]** |

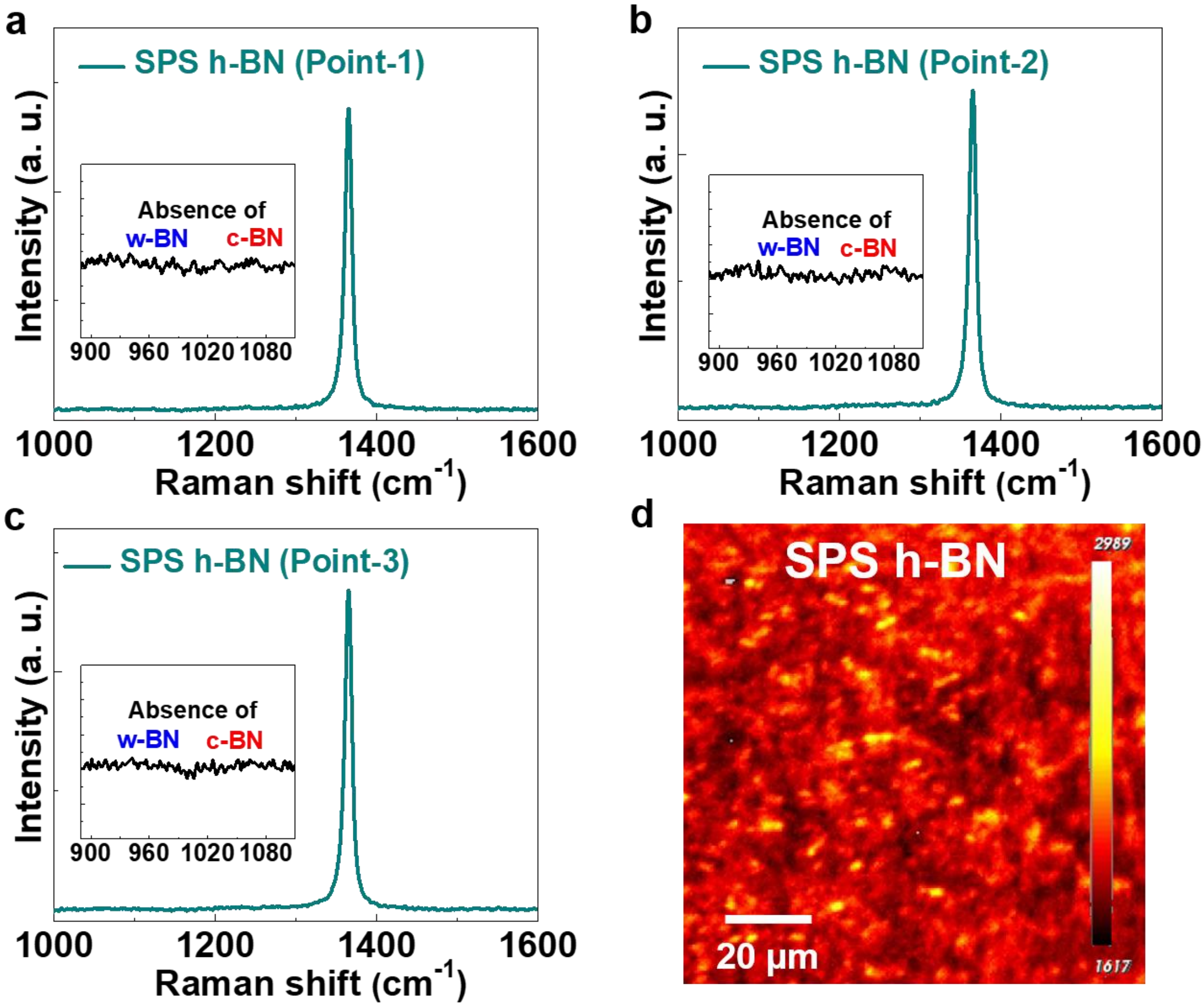

**Fig. S11 Raman spectroscopy of SPS h-BN. a), b), c)** Raman spectra taken at several points of the SPS h-BN disk, showing the sharp $E_{2g}$ Raman peak of h-BN with FWHM of ~12-13 cm$^{-1}$, indicating excellent crystalline quality. **d)** Raman mapping of SPS h-BN surface. Inset shows the search for other phase of BN, for example w-BN (at ~960 cm$^{-1}$) or c-BN (at ~1054 cm$^{-1}$) related peaks, however without any signature.

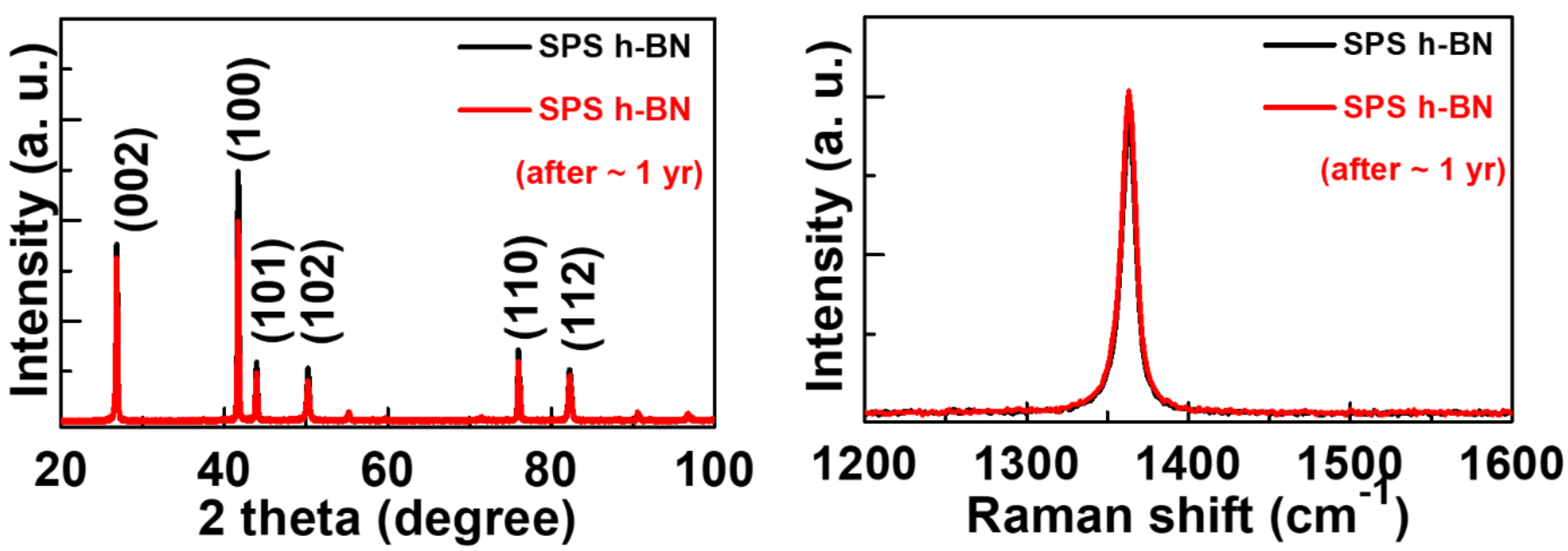

**Fig. S12 Time dependent structural stability test of SPS h-BN**. Once the h-BN is formed via SPS it does not change the structure and remains highly stable over time, without other phases.

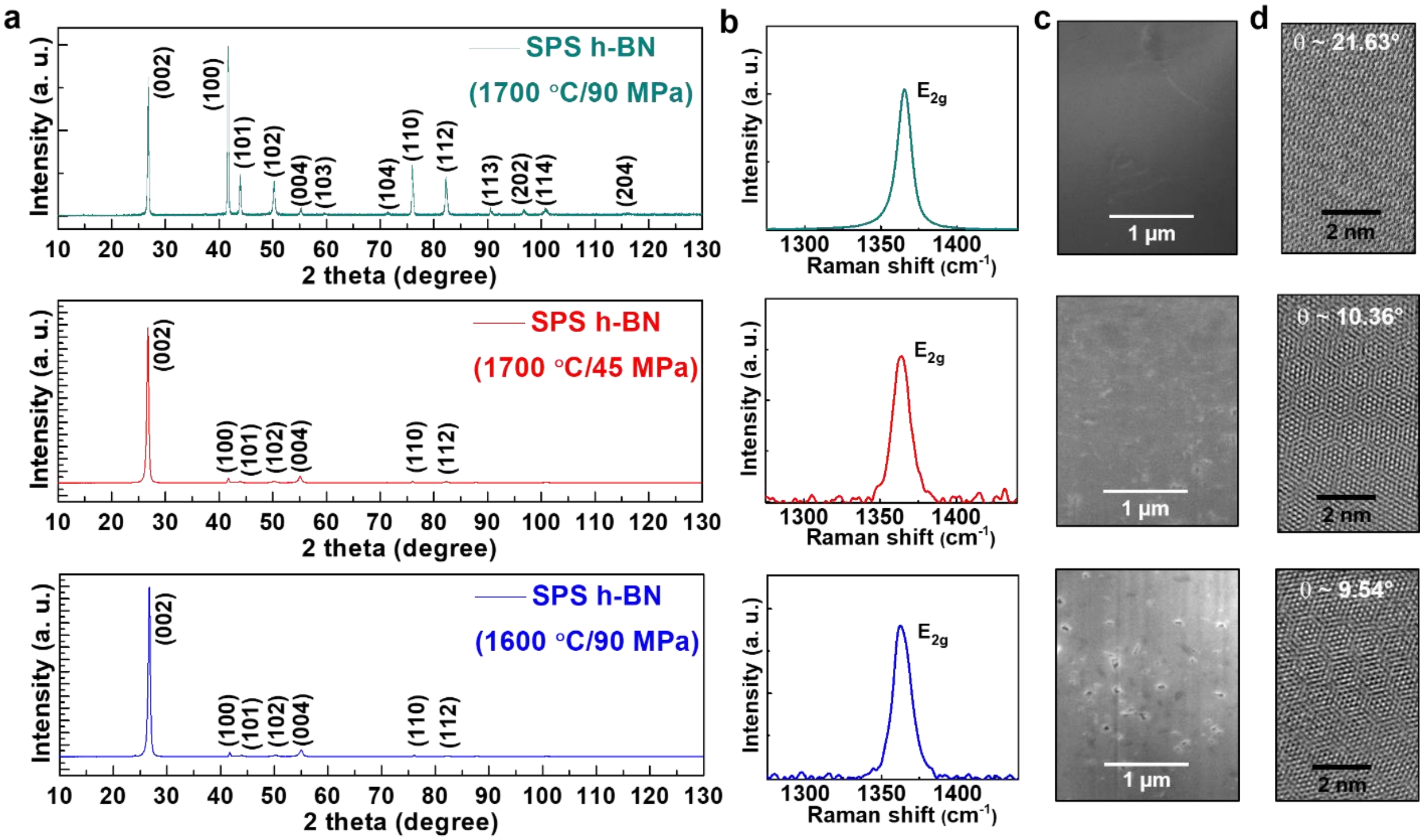

**Fig. S13 Structural characterizations of h-BN sintered at various conditions. a)** XRD, **b)** Raman spectroscopy, **c)** Cross-sectional FESEM, and **d)** Moiré patterns and twist angles for h-BN sintered at 1700 °C/90 MPa, 1700 °C/45 MPa, and 1600 °C/90 MPa for 1 hr.

**Table S4: Comparison of structure and properties of h-BN made at different conditions.**

| SPS condition | Density (gm/cm$^3$) | $I_{(100)}/I_{(002)}$ (from XRD) | Grain size (nm) | Pore-density (%) | Twist angles (°) | Young's modulus (GPa) | Hardness (MPa) | Dielectric constant (at 100 kHz) |
|---|---|---|---|---|---|---|---|---|
| 1700 °C, 90 MPa | 2.05 | 1.2 | ~35.0 | 1.10 | 21.63°, 14.12°, 8.99° | 69.7 ± 3 | 283.3 ± 9 | $\kappa_{\parallel}$ ~9.3, $\kappa_{\perp}$ ~8.2 |
| 1700 °C, 45 MPa | 2.03 | 0.04 | ~25.81 | 1.99 | 10.36° | 47.1 ± 2 | 223 ± 7 | $\kappa_{\parallel}$ ~8.1, $\kappa_{\perp}$ ~5.3 |
| 1600 °C, 90 MPa | 2.02 | 0.03 | ~24.04 | 2.42 | 9.54° | 36.6 ± 3 | 279 ± 11 | $\kappa_{\parallel}$ ~7.3, $\kappa_{\perp}$ ~5.1 |

## Density Functional Theory (DFT) calculations

### Lattice strain dependent calculations

Upon structural optimization, pristine (i.e. non-distorted) h-BN has lattice parameters $a = b =$ 2.50962 Å and $c =$ 6.78547 Å, serving as a model for the CS h-BN material. This structure is shown in which B and N atoms have partial charges of +2.17 and -2.17, respectively (**Fig. S14a**). A second system was generated by distorting the first one by the same percentiles as the SPS h-BN material: compressed by 0.06% along the c axis, and expanded by 0.16% in the ab plan. Both pristine and distorted systems were then compressed uniaxially in each direction. The stiffness values are the same for both models: ~976-1000 GPa (in-plane) and ~44.6 GPa (out-of-plane) (**Fig. S14b**).

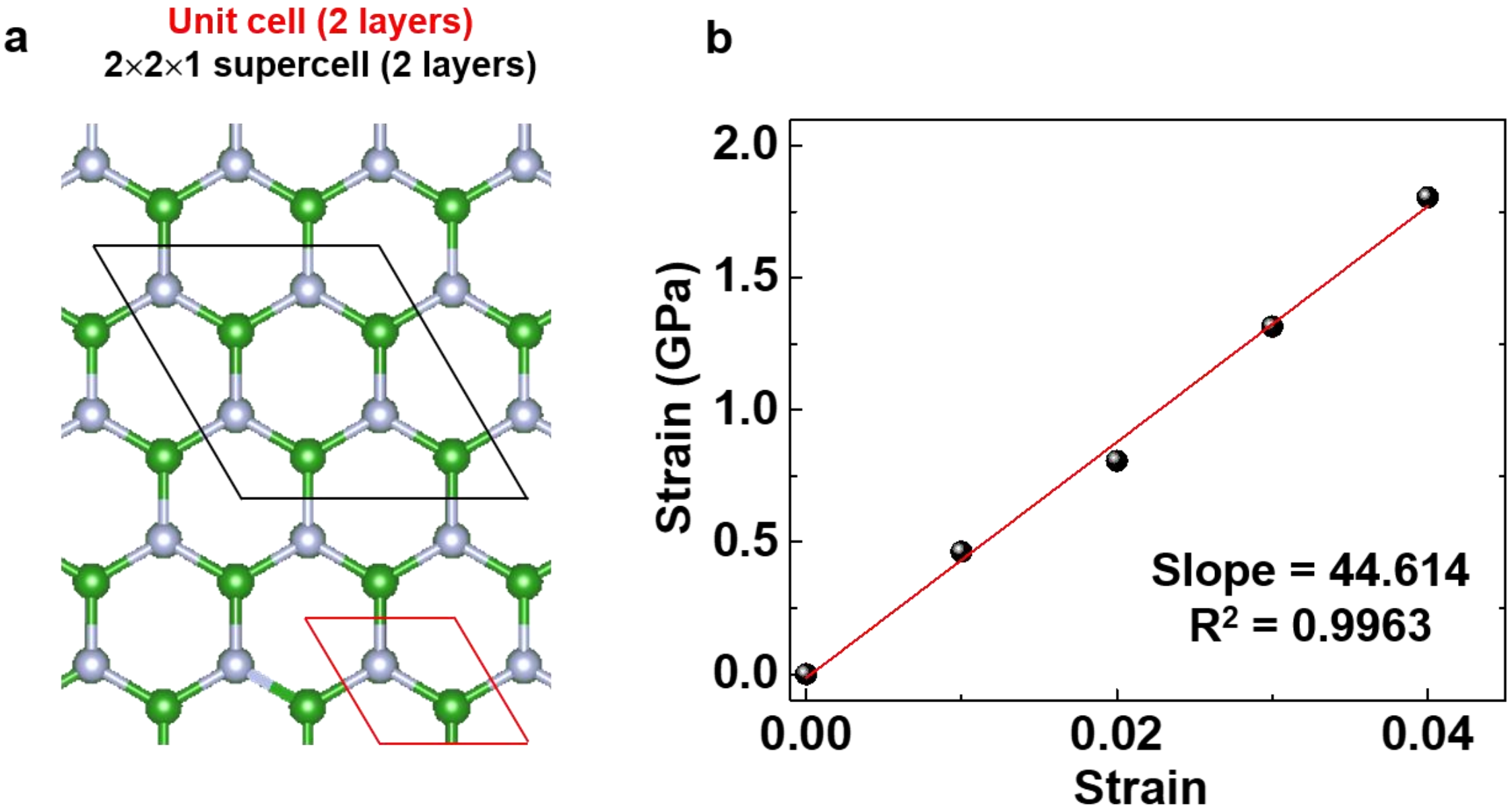

**Fig. S14 Structure and elastic properties of h-BN. a)** Atomic structure with unit cell in red and simulated 2×2×1 supercell in black, each comprised of two layers (green: B atoms, grey: N atoms). **b)** Compressive stress-strain simulation in the out-of-plane direction.

The dielectric constant ($k$) of the pristine system across different frequencies, plotted here in terms of energy (**Fig. S15**). Both the out-of-plane component (Z) and an in-plane component (X, zigzag) of epsilon are shown. The other in-plane direction (Y, armchair) displays the same profile as X; and the distorted BN system has an almost identical behavior as the pristine one. These results are similar to those previously reported in the literature for h-BN.[**S2**] The static dielectric constant, i.e. the value of the dielectric constant at zero energy, is 2.34 along X and 1.64 along Z for the pristine system. Upon distortion of the lattice, these values are increased by 0.04% (in-plane) and 0.09% (out-of-plane), respectively, which is in the same order of magnitude as the changes in the h-BN lattice after SPS. Aiming at investigating ferro- and piezo-electric behaviors, the dipole moment of the pristine and distorted systems were estimated without strain and with 5% compressive strain in each direction separately (zigzag, armchair and out-of-plane). For all eight cases, a total dipole moment of zero was found, suggesting the absence of ferro- and piezo-electric properties in SPS h-BN.

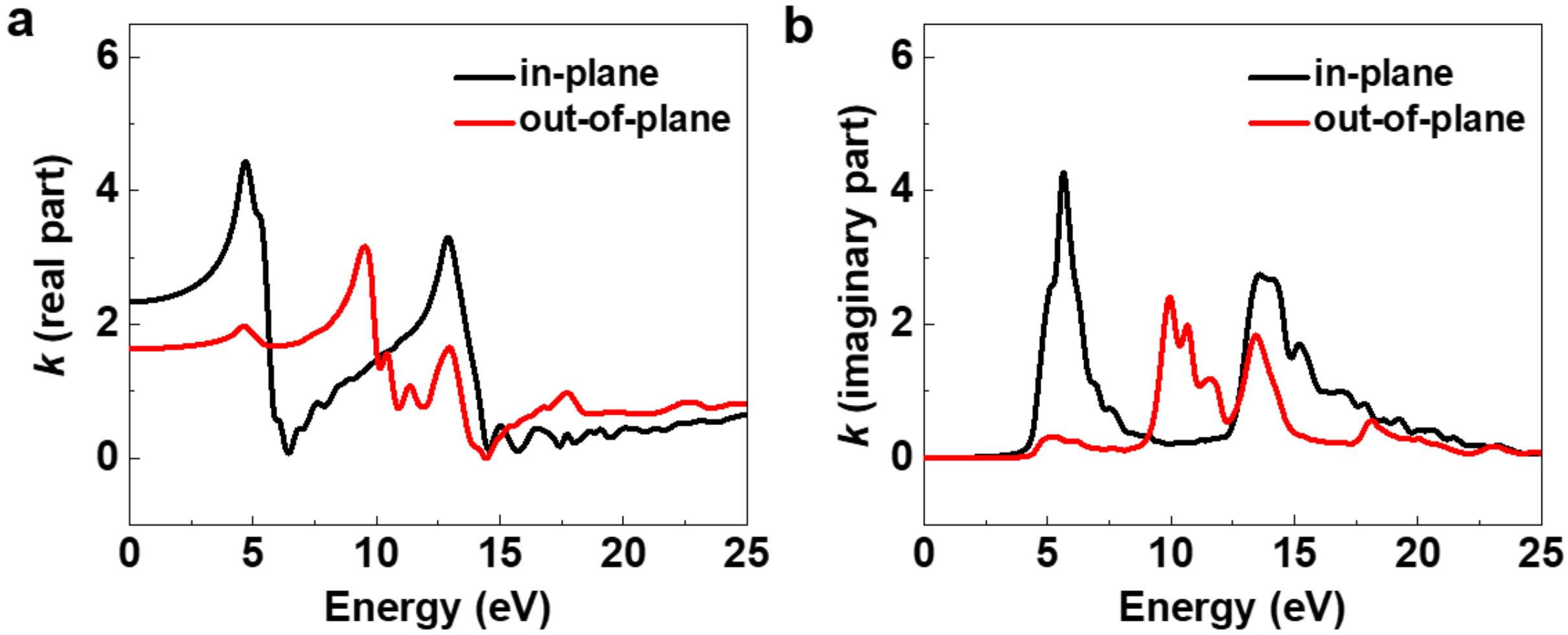

**Fig. S15 Dielectric constant of h-BN.** Calculated **a)** real and **b)** imaginary parts of the dielectric tensor of h-BN along an in-plane (X) and the out-of-plane (Z) directions.

**Twist-angle dependent calculations**

We also estimated the *E* and static out-of-plane *k* of twisted h-BN (**Fig. S16**), and compared them to the pristine system. A twist angle of 21.63° was used (as obtained from HRTEM for our main 1700 °C/90 MPa SPS h-BN sample), which results in a unit cell that is small enough for DFT calculations. However, as seen the twist in h-BN makes a little impact in both *E* and *k*.

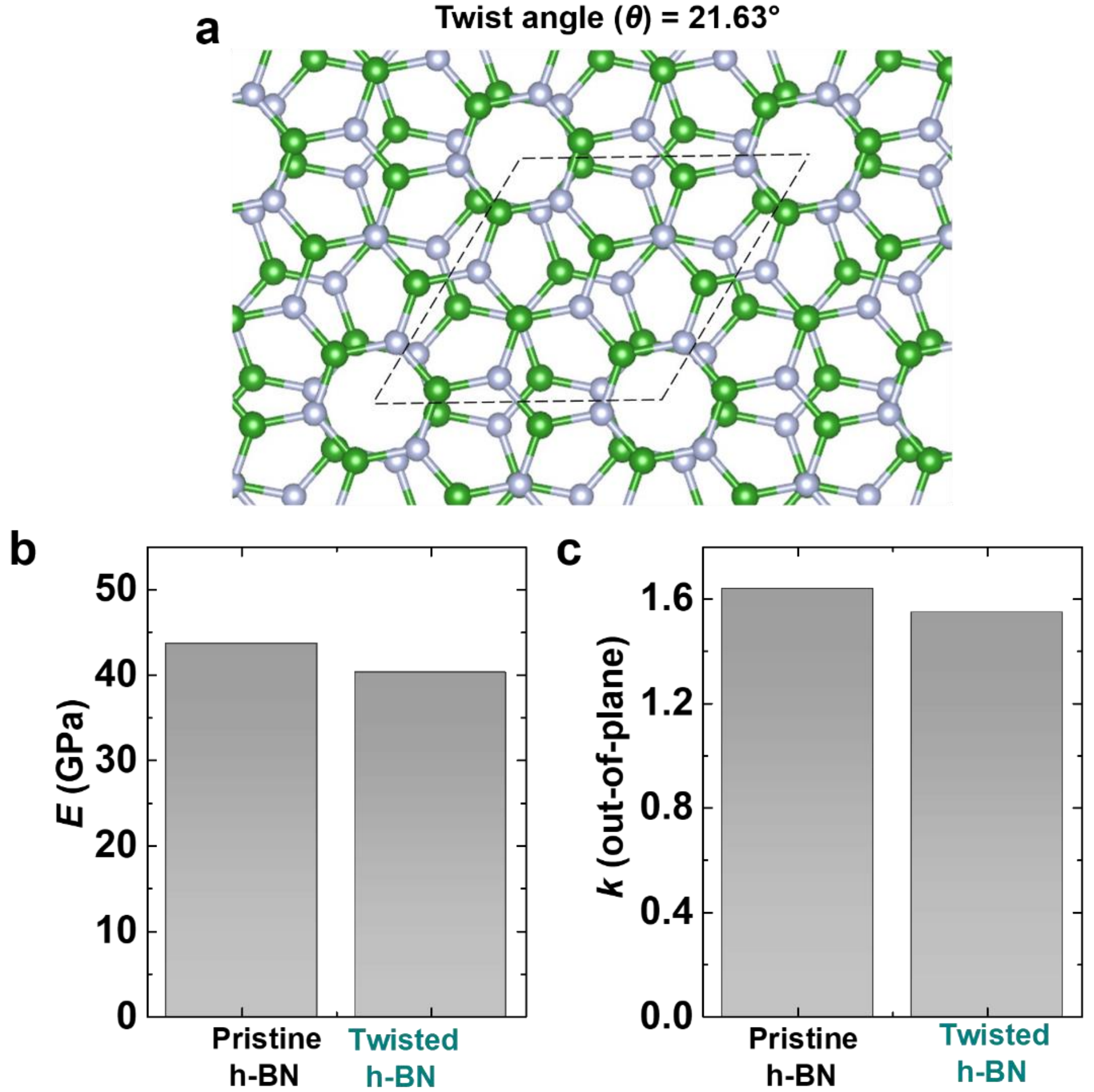

**Fig. S16 Twist dependent properties h-BN.** Calculations based on 21.63° twist angle does not show significant changes in mechanical or dielectric properties.

## Reference

S1. A. Falin, Q. Cai, E. J. G. Santos, D. Scullion, D. Qian, R. Zhang, Z. Yang, S. Huang, K. Watanabe, T. Taniguchi, M. R. Barnett, Y. Chen, R. S. Ruoff, L. H. Li, **Mechanical properties of atomically thin boron nitride and the role of interlayer interactions**, Nature Commun. 8 (2017) 15815.

S2. B. Razieh, S. Valedbagi. **Electronic and optical properties of h-BN nanosheet: A first principles calculation**, Diam. Relat. Mater. 58 (2015) 190-195.